

\magnification=\magstep1

\hsize = 31pc
\vsize = 46pc
\font\grand=cmbx10 at 14.4truept

\def\bs{\baselineskip=14.5pt}


\pageno=0
\def\folio{
\ifnum\pageno<1 \footline{\hfil} \else\number\pageno \fi}

\baselineskip=12pt
\phantom{DIAS}
\rightline{ BONN--HE--93--14 \break}
\rightline{ DIAS--STP--93--02 \break}
\rightline{ hep-th/9304125\break}
\rightline{ revised version}
\vskip 0.5truecm

\bs
\tolerance=8000
\parskip=5pt

\def\pa{\partial}
\def\de{\delta}

\def\G{{\cal G}}
\def\K{{\cal K}}
\def\R{{\cal R}}
\def\W{{\cal W}}
\def\S{{\cal S}}

\def\Gph{{\cal G}_{1\over 2}}
\def\Pph{{\cal P}_{1\over 2}}
\def\Qph{{\cal Q}_{1\over 2}}
\def\la{\langle}
\def\ra{\rangle}

\def\d{{\rm dim\,}}

\def\jhw{{j_{\rm hw}}}

\def\Kg{{{\cal K}_\Gamma}}
\def\gc{{\Gamma_{\rm c}}}
\def\g{{\Gamma}}

\centerline
{\grand On the Completeness of the Set of Classical ${\cal W}$-Algebras}

\centerline{\grand Obtained from DS Reductions}

\vskip 0.6truecm
\centerline{L. Feh\'er\footnote*{
On leave from
Bolyai Institute of Szeged University, H-6720  Szeged, Hungary.}
}
\medskip
\centerline{\it Physikalisches Institut der Universit\"at Bonn}
\centerline{\it Nussallee 12, D-5300 Bonn 1, Germany}
\vskip 0.6truecm
\centerline{L. O'Raifeartaigh, P. Ruelle, I. Tsutsui}
\medskip
\centerline{\it Dublin Institute for Advanced Studies}
\centerline{\it 10 Burlington Road, Dublin 4, Ireland}

\vskip 0.6truecm
\centerline{\bf Abstract}
\vskip 0.2truecm

We clarify the notion of the DS --- generalized Drinfeld-Sokolov ---
reduction approach to classical ${\cal W}$-algebras.
We first strengthen an earlier theorem which showed
that an $sl(2)$ embedding
${\cal S}\subset {\cal G}$ can be associated to
every DS reduction.  We then use the fact that a $\W$-algebra
must have a quasi-primary basis
to derive severe restrictions on
the possible reductions corresponding to a given $sl(2)$ embedding.
In the known DS reductions found to date,
for which the $\W$-algebras are denoted by
${\cal W}_{\cal S}^{\cal G}$-algebras and are called canonical,
the quasi-primary basis corresponds to
the highest weights of the $sl(2)$.
Here we find some examples of noncanonical DS reductions leading to
$\W$-algebras which are direct
products of
${\cal W}_{\cal S}^{\cal G}$-algebras and
`free field' algebras with
conformal weights
$\Delta \in \{0, {1\over 2}, 1\}$.
We also show that if the conformal weights of the generators
of a ${\cal W}$-algebra obtained from DS reduction are nonnegative
$\Delta \geq 0$ (which is
the case for all DS reductions known to date), then the
$\Delta\geq {3\over 2}$ subsectors of the weights are necessarily the
same as in the corresponding ${\cal W}_{\cal S}^{\cal G}$-algebra.
These results are consistent
with an earlier result by Bowcock and Watts on the spectra
of $\W$-algebras derived by different means.
We are led to the conjecture that, up to free fields,
the set of $\W$-algebras
with nonnegative spectra $\Delta\geq 0$
that may be obtained from DS reduction is exhausted
by the canonical ones.

\vfill\eject

\magnification=\magstep1
\hsize = 31pc
\vsize = 46pc

\baselineskip=16pt

\tolerance 2000
\parskip=6pt


\centerline{\bf 1. Introduction}
\medskip

The study of nonlinear extensions of
the Virasoro algebra by conformal primary
fields was initiated by A. B. Zamolodchikov
in the pioneering paper [1].
Such algebras, known as  ${\cal W}$-algebras,
play important r\^ole in two dimensional
conformal field theories, gravity models and integrable systems.
(For detailed reviews, see {\it e.g.} [2,3].)
At least three distinct methods are used in the literature
for constructing ${\cal W}$-algebras.
These can be labelled as direct constructions [1,4,5],
the methods of extracting ${\cal W}$-algebras from conformal
field theories
(the most important of which is the coset construction)
[6,7,8,9,10],
and Hamiltonian reductions of affine Kac-Moody (KM)
algebras to $\W$-algebras [11,12,13].
The Hamiltonian KM reduction method has been intensively pursued
recently both in the classical [14,15,16,17] and in the
quantum framework [18,19,20,21,22], and proved
to be the most productive source of $\W$-algebras so far.
As reviewed in [23], the $\W$-algebras obtained by this method are
symmetry algebras of Toda type field theories, first studied in [24].
(See also [25,26].)

In constructing a reduction of the KM Poisson bracket algebra
to a classical $\W$-algebra
we start by imposing certain {\it first class} constraints on the KM
current, and
consider the ring $\R$ of differential polynomials in the current
which are invariant under the gauge transformations generated by the
first class constraints.
The crucial questions are:

\item{(A)} {\it free generation:}
whether the
ring $\R$ of differential polynomial invariants is freely generated.

\item{(B)} {\it conformal property:}
whether
\itemitem{(b1)} $\R$ contains a gauge invariant Virasoro density.
\itemitem{(b2)} $\R$ has a $\W$-basis.

\noindent
Here by $\W$-basis is meant a basis which consists of a Virasoro
density and fields which are primary with respect to this Virasoro.
See also Section 2 for the notion of {\it classical $\W$-algebra}
used throughout the paper.

These are {\it quite separate} issues and it is easy to construct
examples for which (A) holds but not (B) and (b1) holds but not (A).
(They are of course interrelated since (b2)
obviously requires (A) and (b1).)
Naturally, the answers to {\it both} (A) and (B)
must be positive
for a KM reduction to produce a classical $\W$-algebra.

All Hamiltonian KM reductions that are, to the date of writing,
known to produce a (classical) $\W$-algebra are so called
{\it DS} --- generalized Drinfeld-Sokolov --- reductions.
In a DS reduction one makes the
technical assumption that a certain
mechanism is applicable, whose essence is that a
freely generating basis (not necessarily the $\W$-basis) of $\R$
can be constructed by a gauge fixing procedure,
and the Virasoro element of the $\W$-basis is obtained by
improving the Sugawara formula by adding the derivative of a
current component.
This mechanism is termed the {\it DS mechanism} in this paper
and will be described in some detail.
However, for a generic KM reduction by first class
constraints one cannot expect
the invariant ring $\R$ to
admit a free generating set; in fact, the special
gauge fixing procedure involved in the DS mechanism
is the only known method whereby the
existence of such a basis set can be guaranteed.
The distinguished position of DS reductions among
all Hamiltonian KM reductions derives from
the applicability of this gauge fixing
procedure, which places a
stringent restriction on the nature of the constraints.

An important recent development concerning the Hamiltonian KM reduction
method has been the realization [15,16] that
a DS reduction can be defined to every embedding of the Lie
algebra $sl(2)$ into the simple Lie algebras.
The construction generalizes the standard case appearing
in the construction of KdV type hierarchies by Drinfeld and Sokolov
[11], which corresponds to the principal $sl(2)$ embedding [14].
We call these DS reductions manifestly based on the
$sl(2)$ embeddings the
{\it canonical DS reductions} and the resultant $\W$-algebras the
{\it $\W_\S^\G$-algebras},
where $\G$ is the simple Lie algebra
and $\S\subset \G$ is the $sl(2)$ subalgebra.
One of the salient features of the $\W_\S^\G$-algebras
is that the conformal weights of the
elements in the $\W$-basis are determined
by the $sl(2)$ spins in the decomposition of the
adjoint representation of $\G$ under $\S$,
and the basis
elements are naturally associated to the
highest weight vectors in this decomposition.
Motivated by their natural, group theoretic definition,
and by the fact that at present the canonical DS reductions
are the only KM reductions known to produce $\W$-algebras,
it is expected that
the $\W_\S^\G$-algebras should have an important
r\^ole to play in the classification of $\W$-algebras.

Our main purpose in this paper is to show that the possible
{\it non}canonical DS reductions are {\it severely} restricted.
We do manage to construct some noncanonical DS reductions, but their
$\W$-algebras
turn out to be
direct products of $\W_\S^\G$-algebras and `free field' algebras.
Another purpose is to clarify the notion of
DS reductions, and thus
furnish a framework which could be used in further study
of KM reductions.

We shall present here a stronger version of our
earlier theorem given in [23]
which shows that an $sl(2)$ embedding can
be associated to every DS type KM reduction.
Most considerations in this paper
on general DS reductions will be based
on this crucial structural result.
The source of the inevitable $sl(2)$ structure
given by the theorem  is that
the existence of a $\W$-basis  in $\R$
(more precisely, we shall only need
the existence of a quasi-primary basis for this)
requires the element of $\G$ defining
the improvement term of
the Virasoro density to be the semisimple element
(or \lq defining vector' in the terminology of [27])
of an $sl(2)$ subalgebra.
An immediate consequence of the $sl(2)$ structure is that the
conformal weights $\Delta$ of the elements in
the $\W$-basis
must necessarily be integral or half-integral.
More importantly, since the classification of $sl(2)$ embeddings is
known, the $sl(2)$ theorem reduces the problem of listing all
DS reductions to the problem of
finding the possible different DS reductions that may belong to a given
$sl(2)$ embedding.
The new results obtained in this paper
indicate  that the possible DS reductions
corresponding to a given $sl(2)$ embedding are
extremely restricted.  We shall prove
that, due again to the existence of a $\W$-basis
(or quasi-primary basis),
the dimension of the gauge subalgebra defining
the constraints must be at least half the maximal
dimension allowed by first classness, which is attained in the
canonical DS reduction, and give restrictions
on the position of the gauge subalgebra inside $\G$
with respect to the $sl(2)$ embedding.
We then show that
if the conformal weights of the $\W$-basis
are nonnegative $\Delta \geq 0$,
which is the case for all
$\W$-algebras known to date,
then the sectors $\Delta \geq {3\over2}$
must be the same as those in the corresponding
$\W_\S^\G$-algebra.
This result may be
thought of as complementary (and consistent)
to a result in [31] on the possible conformal spectra of
$\W$-algebras, since our assumptions are different.
(A more detailed comparison between the results of
[31] and our results  can be found in the Discussion.)
Another important result of this paper
is that we shall prove, by providing examples,
the existence of noncanonical DS reductions to $\W$-algebras
where there
occur extra weights $\Delta = 0$, $1\over 2$, $1$ in addition
to the canonical spectrum.
However, in all those noncanoncial
examples the $\W$-algebra turns out to be a direct product of
the $\W_\S^\G$-algebra with trivial `free fields' carrying the
extra weights, and thus it is
essentially equivalent to the $\W_\S^\G$-algebra.

It is clear that the DS reductions form only a special
subset of the
possible conformally invariant Hamiltonian KM reductions,
and it is natural to inquire about the situation in the general case.
This question appears largely unexplored at present,
but the series of examples considered in the Appendix
of this paper gives support to the expectation that in the
general case the ring $\R$ is {\it not freely generated}.
Consider, for example, the \lq $W_n^l$-algebras'
proposed by Polyakov and Bershadsky [28,29]
using KM reductions with
mixed (first class and second class) system of
constraints for $\G = sl(n)$.
An investigation [30]
showed that the invariant ring $\R$ can be defined
similarly to the case of the DS reductions,
but there is no guarantee that
$\R$ is freely generated, since DS gauge fixing
is not applicable,
apart from the cases $W_n^2$ with $n$ odd
which are in fact
equivalent to particular $\W_\S^\G$-algebras.
In other words, in general
the familiar Bershadsky-Polyakov reductions
cannot be expected  to yield
$\W$-algebras in the usual sense of the word, since
they  fail on requirement (A).
Focusing on the particular cases of the $W_{2n}^2$-algebras,
we shall  prove that $\R$ is indeed not freely generated.
{}On this basis we believe that the structure of
the invariant ring $\R$ is in general much more
complicated than in the case of DS type reductions,
which provides the justification
for adopting the applicability of DS gauge fixing as one of
the main assumptions underlying our present study.

This paper is organized as follows.  To clarify the
notion of the DS approach to classical $\W$-algebras
in the more general framework of Hamiltonian KM reductions,
we provide in Section 2 a detailed account of the DS approach
and in particular of the canonical DS
reductions leading to the $\W_\S^\G$-algebras.
Section 3 contains the $sl(2)$ theorem, which
associates an $sl(2)$ embedding to every DS reduction.
Section 4 deals with the restrictions on the possible
DS reductions belonging to the same $sl(2)$ embedding.
Section 5 gives the argument on the
spectrum of the conformal weights, and the new examples of
noncanonical DS reductions.
In section 6 we give
our conclusions, discuss the relationship
of our results with those in [31],
and  point out some open problems.
We conclude with the statement of the conjecture mentioned in the
Abstract and the discussion of some open questions.
There is also an appendix
containing as illustration the
$W_{2n}^2$-reductions
which lead to nonfreely generated invariant rings $\R$.

\vfill\eject
\centerline{\bf 2. Classical $\W$-algebras}
\medskip

By definition, a {\it classical ${\cal W}$-algebra} is a
Poisson bracket algebra built on a finite number of independent fields
$W_a(z)$,
$a=1,\ldots,N$, defined on the circle $S^1$,
according to the following
requirements.
First, the defining Poisson bracket relations are of the
form\footnote{$^1$}{
Conventions:
$
\delta(z - w) :={1\over {2\pi i}}
\sum_{n \in {\bf Z}} z^{n-1} w^{-n}
$
is the delta-function on $S^1$ ($\vert z \vert = \vert w \vert =
1$) for which
$
\oint dz \, f(z) \delta(z - w) = f(w);
$
we use
$\delta^{(i)} (z - w) = ({{d}\over{dz}})^i \delta(z - w)$.
The Virasoro $W_1(z)$ in (2.2) and the KM current $J(z)$ in (2.4)
have their Laurent modes,
$
L_n := i \oint dz \, W_1(z) z^{n +1}
$
and
$
J_n := - i \oint dz \, J(z) z^{n},
$
which fulfil
the standard Virasoro and affine KM algebras
with centre $c$ and $k$, respectively (up to an overall factor
$(-i)$ to be replaced by $1$ upon quantization).
}
$$
\{W_b(z),W_c(w)\}=\sum_i P_{bc}^i
\left(W_1(w),\ldots,W_N(w)\right) \de^{(i)}(z-w),
\eqno(2.1)$$
where the
$P_{bc}^i$ appearing in the {\it finite} sum ($i = 0, 1, 2, \ldots$)
on the right
hand side are {\it differential polynomials}
in the generator fields
$\{W_a\}_{a=1}^N$, with constant terms allowed.
Second, $W_1$ satisfies the Virasoro Poisson bracket algebra,
$$
\{W_1(z),W_1(w)\}=-W'_1(w)\de(z-w) + 2 W_1(w)\de'(z-w) +
{c\over{24\pi}} \de'''(z-w).
\eqno(2.2)
$$
Third, the rest of the
generators $W_a$, $a=2,\ldots, N$, are conformal primary
fields with respect to $W_1$,
$$
\{W_1(z),W_a(w)\}=-W'_a(w)\de(z-w) + \Delta_a W_a(w) \de'(z-w),
\qquad  a=2,\ldots,N.
\eqno(2.3)$$
The classical Virasoro centre $c$ and
the conformal weights $\Delta_a$, $a=2,\ldots,N$, are
(in general complex) numbers.
These constant parameters and the \lq structure polynomials'
$P_{bc}^i$
are restricted by the antisymmetry and the
Jacobi identity of the Poisson bracket.
Two classical $\W$-algebras are regarded to be
equivalent if their
defining relations can be brought to the same form by
a differential polynomial change of basis,
$W_a\rightarrow {\tilde W_a}={\tilde W}_a(W_1,\ldots,W_N)$,
such that the inverse transformation,
${\tilde W_a}\rightarrow W_a=W_a({\tilde W}_1,\ldots,
{\tilde W_N})$
is also given by differential polynomials.

In principle, one can construct classical ${\cal W}$-algebras
by determining the constant parameters
and the structure polynomials directly from the requirements of
antisymmetry and Jacobi identity.  However, in practice
this is hard to carry out systematically, and for this reason
in this paper we are interested in the DS reduction
approach where these requirements are guaranteed by construction.
Moreover, this approach enables us to
quantize the resulting algebras directly through the BRST procedure.

\medskip
\noindent
{\bf 2.1. The DS reduction approach to classical ${\cal W}$-algebras}
\medskip

The general strategy of the DS reduction
approach to constructing classical $\W$-algebras
may be formulated as follows.
Consider a finite dimensional complex simple Lie algebra $\G$ with
the $\rm ad$-invariant, nondegenerate scalar product
$\langle\ ,\ \rangle$.
Denote by $\K$ the space of $\G$-valued
smooth functions on the circle,
${\cal K}:=\{\,J(z)\,\vert\, J(z) \in \G\,\}$,
and let ${\cal K}$ carry the `KM Poisson bracket algebra' given by
$$
\{
\langle \alpha , J(z)\rangle \,,\,\langle \beta , J(w)\rangle \}
=\langle [\alpha,\beta],J(z)\rangle\delta (z-w)
+ K \langle \alpha,\beta \rangle \delta^{\prime}(z-w)\,,
\quad \forall\, \alpha,\beta \in \G,
\eqno(2.4)
$$
where $k = - 2\pi K \neq 0$ is the KM level.
(In other words, the space
$\K$ is the fixed level Poisson
subspace of the dual of the affine KM Lie algebra carrying
the Lie-Poisson bracket.)
We henceforth set the constant $K$ to $1$ for notational
simplicity.
Let us choose a subalgebra $\Gamma\subset \G$, with a
basis $\{\gamma_i\}$
and an element $M\in \G$ in such a way that the following constraints
$$
\phi_i(z)=0\, ,
\qquad\hbox{where}\qquad
\phi_i(z):= \langle\gamma_i\,,\, J(z)-M\rangle\,,
\eqno(2.5)$$
are {\it first class}.
This means that the scalar product $\la\ ,\ \ra$ and the
antisymmetric 2-form $\omega_M$ on $\G$ defined by
$$
\omega_M(\alpha,\beta):=\la M,[\alpha,\beta]\ra,
\qquad \forall\,\alpha,\beta \in \G,
\eqno(2.6)$$
vanish when restricted to $\Gamma$.
The constraint surface, $\K_\Gamma \subset \K$,
defined by (2.5) consists of currents of the form
$$
J(z) = M + j(z),
\qquad j(z) \in \Gamma^\perp\,,
\eqno(2.7)$$
and the first class constraints $\phi_i$ generate gauge
transformations on it,
$$
j\longrightarrow {\rm Ad\,}_{e^{f}}(j):=
e^f(j+M)e^{-f} - M + (e^f)' e^{-f}\,,
\qquad  f(z) \in \Gamma .
\eqno(2.8)$$
We are interested in the {\it gauge invariant differential polynomials}
in $j(z)$ since, as we shall see below,
under certain conditions they furnish a classical
${\cal W}$-algebra.

Let ${\cal R}$ be the set of gauge invariant differential
polynomials in the components of $j(z)$ (with constant terms allowed).
This set is obviously closed with respect to linear combination,
ordinary multiplication and application of $\pa$.
We express this by saying that ${\cal R}$ is a {\it differential ring}.
On the other hand, the induced Poisson bracket
carried by the gauge invariant functions on $\K_\Gamma$
(inherited from the Poisson bracket on $\cal K$)
also closes on $\R$.  Namely, if
$T$, $U\in \R$ one has
$$
\{T(j(z)),U(j(w))\}=\sum_i P_{TU}^i(j(w))\pa^i
\de(z-w)\,,
\eqno(2.9)
$$
where the sum is finite, and $P^i_{TU}\in \R$ because of
the gauge invariance.
This implies that if ${\cal R}$ is
a {\it freely generated} differential ring,
{\it i.e.}, if there exists a basis $\{W_a\}_{a=1}^N \subset \R$
such that any element of $\R$ can be expressed in a unique way
as a differential polynomial in the $W_a$'s,
then the KM Poisson brackets of the basis elements
give an algebra of the form (2.1).
In particular, when it is possible
to find a {\it $\W$-basis} of $\R$ ---
by which we mean such a free generating set
for which (2.2) and (2.3) also hold ---
then we have a classical $\W$-algebra because
the Jacobi identity and antisymmetry are
guaranteed by construction.
Thus, within this approach, our purpose should just be to classify
the KM reductions for which
the invariant ring $\R$ is freely generated
and admits a $\W$-basis.

\medskip

\noindent
{\it 2.1.1. Freely generated ring due to DS gauge}
\smallskip

It is rather obvious that $\R$ is not
freely generated for a generic first class reduction.
For instance, the reductions proposed by Polyakov and Bershadsky
[28,29] aimed at constructing the \lq $W_n^l$-algebras' lead in
general to a nonfreely generated ring $\R$
(see Appendix).
To our knowledge, the only systematic method by which
one can produce free generators for $\R$
relies on the so called DS gauges, the existence of
which places a strong restriction on the reductions.
These gauges may be defined as follows.

\medskip\noindent
{\bf Definition (DS gauge).}
Given a set of first class constraints of type (2.5),
we have a {\it DS gauge} if the following conditions {\rm i) - iii)}
are met:

\item{{\rm i)}}
There exists a diagonalizable element\footnote{$^2$}{A diagonalizable
element defines a grading of
$\G$ by means of its eigenvalues in the adjoint representation,
and is often referred to as a grading operator of $\G$.}
$H\in \G$ such that
$$
[ H,\Gamma ]\subset \Gamma,
\qquad
[ H , M]=-M\,.
\eqno(2.10)
$$
\item{{\rm ii)}}
With a graded linear space $V$ ({\it i.e.} $[H,V]\subset V$)
defining a direct sum decomposition,
$$
\Gamma^\perp = [M,\Gamma] +V\,,
\qquad\hbox{with} \quad V\cap [M,\Gamma]=\{0\}\,,
\eqno(2.11)
$$
one can gauge-fix the constrained current (2.7)
into the form belonging to
the subspace ${\cal C}_V\subset {\cal K}_\Gamma$ given by
$$
{\cal C}_V := \{ J \,\vert\
  J(z) = M + j_{\rm DS}(z) \,,\quad j_{\rm DS}(z)\in V\,\}.
\eqno(2.12)
$$
\item{{\rm iii)}}
The resultant gauge fixed current
$j_{\rm DS}(z) = j_{\rm DS}(j(z))$,
in which  the gauge orbit passing through
$j(z)\in {\cal K}_\Gamma$ meets the global gauge section
${\cal C}_V$,
is given by a {\it differential polynomial in the
original current $j(z)$.}

\medskip
Condition i) requires a special element $H$ whose
adjoint action ${\rm ad}_H$ maps $\Gamma$ into itself and
with respect to which $M$ is an eigenvector with nonzero
eigenvalue, $[H,M]=\lambda M$ (for later convenience
we have scaled $H$ so that $\lambda = -1$).
Note that it is not always possible to find such an $H$ for a given
pair $(\Gamma,M)$,
even if we take into account that $M$ can be redefined
by $M\to M+m$, $m\in \Gamma^\perp$, which
does not affect the constraints (2.5).
The main requirement given by ii), iii) is that
${\cal C}_V$ is a global gauge section of (2.8)
such that the components of
the gauge fixed current $j_{\rm DS}$,
when considered as functions on ${\cal K}_\Gamma$,
are elements of ${\cal R}$.
In particular, if $\Gamma$ consists of nilpotent elements of $\G$
then iii) is implied by the
stronger and more practical requirement
\item{{${{\rm iii)}}'$}}
The gauge-fixing equation corresponding
to the gauge section ${\cal C}_V$,
$$
j\rightarrow {\rm Ad}_{e^f}\,j=
e^{f}j e^{-f}
+(e^{f}Me^{-f}-M)+
(e^{f})' e^{-f}=
j_{\rm DS}\,,
\eqno(2.13{\rm a})
$$
where
$$
j(z)\in \Gamma^\perp\,,
\qquad
f(z)\in \Gamma\,,
\qquad
j_{\rm DS}(z)\in V\,,
\eqno(2.13{\rm b})
$$
has a {\it differential polynomial
solution} $f(z)=f(j(z))$.

\noindent
In all known examples for which
a DS gauge exists, $\Gamma$
actually consists of nilpotent elements
and one has property ${{\rm iii)}}'$.  This will include all the
examples given in this paper.

\medskip

When a DS gauge is available,
we call the procedure by which the general first
class constrained current is transformed to such a gauge,
{\it i.e.}, whereby eq.~(2.13) is solved,
the {\it DS gauge fixing procedure} [11,23].
Note that in principle we need not require that the solution
be unique for the gauge transformation $e^f$
though it is actually unique in all known examples.
Like for any gauge invariant function,
for any  $P(j)\in\R$ we have
$$
P(j)=P({\rm Ad}_{e^{-f}}\, j_{\rm DS})
=P(j_{\rm DS}(j))
\eqno(2.14)
$$
by inverting (2.13a).
The point is that by using iii) this equation now implies
that the components of $j_{\rm DS}(j)$ form a
generating set for $\R$.
Furthermore,
these generators of $\R$ are independent since they reduce
to independent current components in the DS gauge, {\it i.e.},
we have
$$
j_{\rm DS}(j(z))=j(z) \qquad {\rm on} \quad {\cal C}_V \,,
\eqno(2.15)$$
which follows directly from the notion of gauge fixing.
In conclusion, we see that {\it if a DS gauge exists then
$\R$ is freely generated, and
a basis is given by the components of $j_{\rm DS}(j)$}.

Clearly, the number of components of the gauge fixed current
$j_{\rm DS}$ should  be $\d \G -2\d\Gamma$, and this implies
by (2.11-12) that the {\it nondegeneracy} condition,
$$
{\rm Ker\,}\left({\rm ad\,}_M\right)\cap \Gamma =\{\, 0 \,\}\,,
\eqno(2.16)$$
is a {\it necessary} condition for DS gauge fixing.
On the other hand, we can provide  a reasonably
simple {\it sufficient} condition for DS
gauge fixing as follows [23].
Choose a graded subspace $\Theta\subset\G$ which is dual to
$\Gamma$ with respect to the 2-form $\omega_M$, and
define $V$ in (2.11) to be
the space orthogonal to both $\g$ and $\Theta$,
$$
V:=\Theta^\perp\cap\Gamma^\perp\,.
\eqno(2.17)$$
In other words, add the gauge fixing conditions
$$
\chi_k(z):=\la \theta_k,J(z)-M\ra =0\,,
\qquad \theta_k\in \Theta\,,
\eqno(2.18)$$
to the first class constraints (2.5).
If, in addition to the nondegeneracy condition (2.16),
one has
$$
[\Theta\,,\,\Gamma]_{\geq 1}\subset \Gamma\,,
\eqno(2.19)$$
where the subscript refers to the grading defined by $H$,
and if $\Gamma$ consists of {\it nilpotent} elements of $\G$,
then by using $V$ in (2.17) one indeed obtains a DS gauge.
(We refer the reader to [23] for a detailed description of
the recursive DS gauge fixing procedure based on this sufficient
condition.)
A somewhat stronger sufficient condition for DS gauge fixing
is furnished by complementing (2.16) with the condition
$$
\Gamma^\perp \subset \G_{>-1}\,.
\eqno(2.20)$$
Equations (2.16) and (2.20) together imply
$$
\G_{\geq 1}\subset \Gamma\subset \G_{>0}\,,
\eqno(2.21)$$
which ensures (2.19) and that $\Gamma$ consists of nilpotent elements.

We also note the following further consequences of the
definition of a DS gauge.
First, because of (2.15),
the components of $j_{\rm DS}(j(z))$, defined by means
of a basis of $V$, contain the corresponding components of $j(z)$ in
their linear terms.
Second, for the very same reason, the induced KM Poisson
bracket algebra of the components of $j_{\rm DS}(j(z))$
can be identified with the Dirac bracket algebra carried
by the components of the gauge fixed current,
where the second class constraints
defining the Dirac bracket are
given by combining (2.5) and (2.18) together [14].

\medskip
\noindent
{\it 2.1.2. The form of the Virasoro density and the DS
mechanism}
\smallskip

Having assumed the existence of a DS gauge using the
grading operator $H\in \G$, next we have to ensure
that the polynomial Poisson bracket algebra carried by $\R$
contains the Virasoro subalgebra.  For this we shall consider
the following density,
$$
L_H:= {1\over 2}\la J,J\ra -\la H,J'\ra\,.
\eqno(2.22)$$
Indeed, one can easily check that this
defines a {\it gauge invariant} Virasoro density, {\it i.e.},
it not only fulfils the Virasoro algebra
but also is an element of $\R$, provided that
in addition to (2.10) one has
$$
H\in \Gamma^\perp\,.
\eqno(2.23)$$
Of course, the relations (2.10) and (2.23)
also imply that the conformal action
generated (for $\delta_{f} z=-f(z)$)
by the charge
$Q_f=\oint dz f(z) L_H(z)$ on $\K$,
$$
\delta_{f} J:=-\{ Q_f\,,\, J\} =f J' + f'(J + [H,J]) + f'' H\,,
\eqno(2.24)$$
induces a conformal action on the space of gauge orbits in
$\K_\Gamma$.
We note that the coefficient of the term
$\la H,J'\ra$ in $L_H$ rescales according to the choice of
$\lambda$ in $[H, M] = \lambda M$; the value $-1$
in (2.22) is adjusted for our choice $\lambda = -1$.
Note also that (2.23) is a rather mild
additional assumption to the existence of a DS gauge,
since in the examples when DS gauges exist $\Gamma$
is usually a strictly
triangular subalgebra of $\G$
and (2.23) is automatic for the Cartan element $H$.

Based on the construction we described above,
and motivated by the canonical DS reductions which we will
recall in the next
section, we are interested in reductions
of KM Poisson bracket algebras to classical $\W$-algebras
through the following mechanism:

\item{i)}
 {\it The first class constraints $(2.5)$ admit a DS gauge
 with respect to a grading operator $H$.}

\item{ii)}
  {\it There exists a $\W$-basis in the invariant ring
  $\R$ with respect to the
  Virasoro density} $W_1:=L_H$.

\noindent
These assumptions imply the existence of a basis of
$\R$ yielding a classical $\W$-algebra in the sense of
eqs.~(2.1-3), and we believe
that they are not much stronger than requiring just
this to be the case.
(To the date of writing, we have no counterexample.)
By definition, in this paper
we call a KM reduction defined by first
class constraints of type (2.5) a {\it DS reduction}
if the above {\it DS mechanism} i), ii) is applicable.

Before describing the canonical DS reductions where this mechanism
is at work, and whose uniqueness is the main question addressed later,
we wish to mention another consequence of the assumptions.
Namely, we observe from (2.24) that if $f''=0$ then the infinitesimal
conformal transformation generated by $Q_f$ leaves the DS gauge
fixed current form invariant, and we have
$$
\delta_{f}\, j_{\rm DS} =f {j'}_{\rm DS} +f'(j_{\rm DS} +
[H,j_{\rm DS}])\,,
\qquad\hbox{for}\qquad
f''=0.
\eqno(2.25)$$
Since $f''=0$ holds for the infinitesimal
scale transformation for which $f(z)\sim z$,
we see from (2.25) that the components of $j_{\rm DS}(j)$
have definite {\it scale dimensions} given by shifting
the grades of the corresponding basis elements of $V$ in
(2.11) by $+1$.

\medskip
\noindent
{\bf 2.2. The canonical DS reductions and
the ${\cal W}_{\cal S}^{\cal G}$-algebras}
\medskip

The DS mechanism works in the {\it canonical DS reductions}
which are associated to the $sl(2)$ embeddings
in $\G$ in the following way.
Let $\S= \{ M_-,M_0,M_+ \} \subset \G$ be
an $sl(2)$ subalgebra
with standard commutation relations
$$
[M_0,M_\pm]=\pm M_\pm\,,\qquad[M_+,M_-]=2M_0.
\eqno(2.26)$$
Consider the grading of $\G$ defined by the
eigenvalues of ${\rm ad\,}_{M_0}=[M_0,\ ]$,
$$
\G=\sum_{m}\G_m,
\qquad {\rm where} \qquad [M_0,X]=mX, \quad \forall\,X\in \G_m.
\eqno(2.27)$$
Choose a subspace $\Pph\subset\Gph$ for which
$$
\omega_{M_-}(\Pph,\Pph)=\{ 0\},
\qquad
{\rm dim\,}\Pph={1\over 2}{\rm dim\,}\Gph,
\eqno(2.28{\rm a})$$
and define the {\it canonical subalgebra} $\Gamma_{\rm c}$ by
$$
\Gamma_{\rm c} :=\Pph +\G_{\geq 1}\,.
\eqno(2.28{\rm b})$$
The canonical first class constraints
are  obtained from
(2.5) by taking $\Gamma:=\Gamma_{\rm c}$ and $M:=M_-$, and thus
the constrained current takes the form
$$
J(z)=M_-+j_{\rm c}(z)\,,
\quad j_{\rm c}(z)\in \Gamma_{\rm c}^\perp
\qquad\hbox{with}\qquad
\Gamma_{\rm c}^\perp=[M_-,\Pph]+{\cal G}_{\geq 0}\,.
\eqno(2.29)$$
Note that with respect to $M:=M_-$ the dimension of
$\Gamma=\Gamma_{\rm c}$ is the maximal one allowed  by the
first classness of the constraints and the nondegeneracy
condition (2.16).
It is also easy to check that
DS gauges are available by using $H:=M_0$ as the
grading operator and that $L_{M_0}\in\R$.

The $\W$-basis of $\R$ is constructed
by means of the `highest weight gauge' [14],
which is the particular DS gauge obtained by taking
$$
V:={\rm Ker\,}({\rm ad\,}_{M_+})
\eqno(2.30)$$
in the direct sum decomposition of type (2.11).
For this, we first fix  a basis
$\{\, Y_{l,n}\,\}\subset {\rm Ker\,}({\rm ad\,}_{M_+})$
of highest weight vectors,
$$
[M_0\,,\,Y_{l,n}]=l Y_{l,n}\,,
\qquad
Y_{1,1}:=M_+/\la M_-,M_+\ra\,,
\eqno(2.31)$$
where $n$ is a multiplicity index and $\la M_-,Y_{l,n}\ra=0$
for $Y_{l,n}\neq Y_{1,1}$.
We then write the current
resulting from the gauge fixing,
$j_{\rm hw}(j_{\rm c}(z))$,
in the form
$$
j_{\rm hw}(j_{\rm c}(z))=\sum_{l,n} W_{l,n}(j_{\rm c}(z))\, Y_{l,n}\,,
\eqno(2.32)$$
and $\{\,W_{l,n}(j_{\rm c})\,\}\subset \R$ is a basis of the
invariant ring.
It turns out that, except $W_{1,1}$, $W_{l,n}$ is a
primary field of weight $(l+1)$ with respect to $L_{M_0}$ [15,23].
The Virasoro density $W_1:=L_{M_0}$ given by
$$
W_1(j_{\rm c})={1\over 2}\la M_- + j_{\rm c}\,,\,M_-+j_{\rm c}\ra -
\la M_0, j'_c\ra\,,
\eqno(2.33{\rm a})$$
can be rewritten as
$$
W_1(j_{\rm c})={1\over 2}
\la j_{\rm hw}^{\rm sing}(j_{\rm c}), j_{\rm hw}^{\rm sing}(j_{\rm c})
\ra + W_{1,1}(j_{\rm c})\,,
\eqno(2.33{\rm b})$$
where
$$
j_{\rm hw}^{\rm sing}(j_{\rm c}):=\sum_n W_{0,n}(j_{\rm c})\, Y_{0,n}
\eqno(2.33{\rm c})$$
is the $sl(2)$ singlet part of $j_{\rm hw}(j_{\rm c})$.
The singlet components $W_{0,n}(j_{\rm c})$ generate
a KM algebra under the induced Poisson bracket,
and the first term in (2.33b) is just the corresponding
Sugawara formula.
The second term $W_{1,1}(j_{\rm c})$ in (2.33b) is another
Virasoro density, that commutes with the singlet Sugawara density.
We see from the above  that the required
$\W$-basis, $\{\,W_a\,\}_{a=1}^N\subset \R$,
$N=\d {\rm Ker\,}({\rm ad\,}_{M_+})$, can be obtained
from the basis  $\{\,W_{l,n}\,\}\subset \R$
by exchanging $W_{1,1}$ with $W_1$, and calling the
rest of the basis elements $W_2, \ldots, W_N$.
The resulting classical $\W$-algebra is called the
{\it $\W_\S^\G$-algebra}.
By the remark given at the end of the subsection 2.1.1.,
the $\W_\S^\G$-algebra can be interpreted also as the
Dirac bracket algebra carried by the
components of the current in the highest weight gauge
(after the aforementioned change of basis is made).

It is worth stressing that, apart from those $sl(2)$
embeddings for which there are no singlets
in the adjoint of $\G$,
the $\W_\S^\G$-algebra contains
the singlet KM subalgebra
generating the group of
canonical transformations:
$$
j_{\rm hw}(z)\longrightarrow
e^{\alpha^i(z) Y_{0,i}} j_{\rm hw}(z) e^{-\alpha^i(z) Y_{0,i}}
+(e^{\alpha^i(z) Y_{0,i}})' e^{-\alpha^i(z) Y_{0,i}}\,.
\eqno(2.34)$$
It follows that the generators of the $\W_\S^\G$-algebra given
by the nonsinglet components of $j_{\rm hw}$
fall into representations of
the Lie algebra of the singlets, and that
one can further reduce the $\W_\S^\G$-algebra
by using this sub-KM symmetry, {\it i.e.}, by
putting constraints on the singlet components of $j_{\rm hw}$.
However, such `secondary reductions' do not in general lead to new
$\W$-algebras based on independent fields
according to the requirements (2.1-3) (see also the Appendix).

Having our hands on the above rather nice examples,
it appears natural to ask how close they are to an exhaustive
set of $\W$-algebras that can be obtained through the DS mechanism.
This question will be even more natural after establishing in the
next section that in a certain sense there is indeed
an $sl(2)$ embedding behind any $\W$-algebra obtained in this way.

\vfill\eject
\centerline{\bf 3. The existence of an $sl(2)$ structure}
\medskip

In this section we prove a theorem, which allows
one to associate an $sl(2)$ embedding to every
reduction yielding a $\W$-algebra by means of the DS mechanism
reviewed in the previous section.
More precisely, our assumption will be that $\R$
is freely generated due to the existence of a DS
gauge and possesses a quasi-primary basis with
respect to $L_H$.  We shall then conclude that $H\in \G$
must belong to an $sl(2)$ subalgebra.
This immediately implies that {\it the conformal
weights must be either
integral or half-integral in every $\W$-algebra
arising in this way.}
This theorem is a stronger version of the previous result
in [23],
and the rest of the paper is devoted to uncovering its implications.
To make the proof as clear as possible
we shall proceed through two preliminary lemmas.

Consider conformally invariant first
class constraints described by a triple $(\Gamma, M,H)$.
The form of the constrained current is given by
eq.~(2.7) and $L_H$ (2.22) defines an element of
the invariant ring $\R$.
For any vector field
$f(z){d\over{d z}}\in {\rm diff\,}S^1$,
the infinitesimal conformal transformation
$\delta_{f} J$ is generated by the charge $Q_{f}$
according to (2.24).
This conformal action preserves the constraint surface
${\cal K}_\Gamma\subset {\cal K}$, and we have
$$
\delta_{f}j=fj'+f'(j+[H,j])+f''H .
\eqno(3.1)$$
Consider now the subspace of special configurations,
 ${\cal C}_0\subset {\cal K}_\Gamma$, given by
$$
{\cal C}_0:=\{\, J\,\vert\, J(z)=M+ h(z) H\, \}\, .
\eqno(3.2)$$
This subspace ${\cal C}_0$ is
invariant under conformal transformations,
and the field $h(z)$ transforms according to
$$
\delta_{f} h=fh'+f'h + f''\,.
\eqno(3.3)$$
By definition, $U(z)$ is called a {\it quasi-primary} field
of scale dimension $\Delta$ (which is the
conformal weight if $U(z)$ is primary) if it transforms as
$$
\delta_{f} U=fU'+ \Delta f' U
\eqno(3.4)$$
under the {\it M\"obius subgroup} of the conformal group
generated by the vector fields
with $f'''=0$.  As far as scale dimension $\Delta$
is concerned, it can be
defined even for a non-quasi-primary field $U(z)$
if it satisfies (3.4) for $f''=0$.  For example,
the field $h(z)$ is not quasi-primary
but has scale dimension $1$.
We then have the following statement.

\medskip\noindent
{\bf Lemma 1.}
{\it There is no quasi-primary differential polynomial
$p(j)$ on $\K_\Gamma$ whose restriction to
${\cal C}_0$ (3.2) satisfies
$$
p(j) \vert_{{\cal C}_0}=Ah
\eqno(3.5)$$
with a nonzero constant $A$.}

\medskip\noindent{\it Proof.}
Since ${\cal C}_0$ is invariant under conformal transformations
and $Ah$ is not quasi-primary for $A\neq 0$, we see that
a differential polynomial $p(j)$ satisfying (3.5) cannot be
quasi-primary. {\it Q.E.D.}

\medskip
On the other hand, we also have the following statement.

\medskip\noindent
{\bf Lemma 2.}
{\it Suppose the constraints admit a DS gauge
for which the complementary space
$V$ in $(2.11)$ is graded by $H$, and
$$
H\notin [M,\Gamma]\,.
\eqno(3.6)$$
Then there exists a gauge invariant differential
polynomial $P^H(j)\in \R$ whose
restriction to ${\cal C}_0$ $(3.2)$
is proportional to the field $h$,
$$
P^H(j) \vert_{{\cal C}_0}=Ah,
\eqno(3.7)$$
where $A$ is a nonzero constant.}

\medskip\noindent{\it Proof.}
Let $j_{\rm DS}(j)\in V$ be the
gauge transform of the general current
$j\in \K_\Gamma$ to the DS gauge.
Recall that the components of $j_{\rm DS}(j)$
generate $\R$ and have scale dimensions
given by shifting the grades of the
corresponding basis elements of $V$
by $1$.
Recall also that the components of $j_{\rm DS}(j)$
contain the corresponding components of $j$.
{}From these facts and (3.6) we see that
the restriction of $\la H, j_{\rm DS}(j)\ra$
to ${\cal C}_0$ (3.2) contains a term
proportional to $h$ and has
scale dimension $1$.
Clearly, we can thus take
$P^H(j):= \la H, j_{\rm DS}(j)\ra$ to be the required
element of $\R$. {\it Q.E.D.}

\medskip
The $sl(2)$ theorem will result by combining the statements
of the two lemmas.
The theorem uses the notion of a quasi-primary basis.
By definition, the basis $\{W_a\}_{a=1}^N\subset \R$ is a
{\it quasi-primary basis} if the basis elements are
quasi-primary fields with respect to the given Virasoro density.
Clearly, a $\W$-basis is a quasi-primary basis.

\medskip\noindent
{\bf Theorem.}
{\it Suppose that the conformally invariant first class
constraints described by $(\Gamma,M,H)$
admit a DS gauge with respect to the grading operator $H$.
Suppose furthermore that there exists a quasi-primary
basis of gauge invariant differential polynomials
$\{ W_a\}_{a=1}^N\subset \R$
$(N=\d\G-2\d\Gamma)$ with respect to $L_H$.
Then there exists an element $M_+\in \Gamma$
such that the standard $sl(2)$ commutation relations $(2.26)$
hold with $M_-:=M$ and $M_0:=H$.}

\medskip\noindent{\it Proof.}
Suppose that we have (3.6).  Then by Lemma 2 we have
an element $P^H(j) \in {\cal R}$ whose restriction to
${\cal C}_0$ has the property (3.7).  On the other
hand, since we  assumed a
quasi-primary basis in ${\cal R}$
we can express $P^H(j)$ as a differential polynomial in the basis,
$$
P^H(j)=P(W_1(j),W_2(j),\ldots,W_N(j)).
\eqno(3.8)
$$
When restricted to
${\cal C}_0$ the $W_a(j)$'s in the r.h.s.~of (3.8)
either vanish or become
quasi-primary differential polynomials in $h$.  However, due to
Lemma 1,
none of the nonvanishing ones
can contain a term proportional to $h$ and hence the r.h.s.~of
(3.8) does not reduce to the expression $Ah$.
Since this contradicts (3.7),
we conlude that (3.6) cannot hold.
Thus there must
exist an element
$\gamma\in \Gamma$ such that
$H=[M,\gamma]$. Decomposing $\gamma$ into a grade 1 part and the
rest, $\gamma = \gamma_1 + \gamma_{\neq 1}$, we obtain
$[M,\gamma_{\neq 1}]=0$ on account of the grading. The
nondegeneracy condition (2.16)
then implies $\gamma_{\neq 1}=0$. (The
element $\gamma_{\neq 1}$ must be in $\Gamma$, since $\Gamma$ is
assumed to be graded (2.10).) Thus $\gamma$ has grade 1 and we have
$$
H = [M, \gamma], \qquad {\rm and} \qquad [H, \gamma] = \gamma.
\eqno(3.9)
$$
Combining (3.9) with $[H, M] = - M$ in (2.10), we find that
the set $\{M, H, \gamma\}$ forms the required $sl(2)$
subalgebra ${\cal S} = \{M_-, M_0, M_+\}$ of (2.26).  {\it Q.E.D.}

\medskip

Since a $\W$-basis is necessarily a quasi-primary basis,
the Theorem says that one can always find an $sl(2)$ subalgebra
by the set
given above for any ${\cal W}$-algebra obtained from DS
reduction.
This suggests that the
${\cal W}_{\cal S}^{\cal G}$-algebras, which are manifestly
based on the $sl(2)$ subalgebras of ${\cal G}$,
are in fact \lq natural' in the context of the DS reduction approach.
This in turn leads us to
the question as to whether the
${\cal W}_{\cal S}^{\cal G}$-algebras are the
only classical $\W$-algebras that may be obtained
from DS reduction.
We shall try to answer
this question in the rest of the paper.

\vfill\eject
\centerline
{\bf 4. Restrictions on $\Gamma$ for a given $sl(2)$ embedding}
\medskip

In Section 3 we found that there exists an $sl(2)$ embedding,
$\S=\{M_- = M, M_0 = H, M_+ \in \Gamma \}$,
to any system of conformally invariant first class constraints
given by a triple $(\Gamma, M,H)$ for which $\R$ is freely
generated due to the existence of a DS gauge and possesses
a quasi-primary basis with respect to $L_H$.
This result reduces the problem of listing all DS reductions
to the problem of finding all allowed $\Gamma$'s for given
$sl(2)$ embeddings in $\G$
(whose classification is known),
such that the triple $(\Gamma, M_-,M_0)$ leads to
a ${\cal W}$-algebra.
In this section we shall see
that the same requirement used earlier to uncover the
$sl(2)$ structure,
{\it i.e.}, that there exists a quasi-primary basis in $\R$,
also gives considerably strong restrictions on
the allowed $\Gamma$.
(As in the previous section we only need to
require the existence of a quasi-primary basis in $\R$
rather than a $\W$-basis.)
In particular,  we shall prove that $\Gamma$
must satisfy a certain number of inequalities
on the dimensions of its graded subspaces.  The
simplest ones among them are
$$\d \Gamma_q \geq {1 \over 2} \d \G_q,
\qquad \forall\, q\geq 1,
\eqno(4.1)
$$
which imply that $\Gamma_{\geq 1}$ must be
at least half as large as $(\gc)_{\geq 1}$.
To derive (4.1), we shall use a two-step DS gauge fixing
procedure based on a semi-direct sum
structure of the gauge subalgebra
$\Gamma$.  We shall examine the existence of a quasi-primary basis
by asking if the DS gauge fixed current can be expressed as a
differential polynomial in such a basis.
For this purpose it will be convenient to expand every element of $\R$
as a differential polynomial in the partially gauge fixed current
provided by the first step of the
gauge fixing, described in Section 4.1.
By inspecting the linear term in the expansion of the fully DS
gauge fixed current and requiring that it be compatible
with the existence of a quasi-primary basis,
in Section 4.2 we shall prove a proposition from which the
inequalities in (4.1) follow.

\medskip
\noindent
{\bf 4.1. Gauge fixing with respect to $M_+$}
\medskip

In order to implement the first step of the
two-step gauge fixing for
the system given by the triple $(\Gamma, M_-,M_0)$,
let us write
the first class constrained current
$J(z)\in {\cal K}_\Gamma$ in the form:
$$
J(z)=M_- + h(z) M_0 +j_+(z)M_+ +t(z)\,,
\quad\hbox{with}\quad
t(z)\in \Gamma^\perp\cap \S^\perp\,.
\eqno(4.2)$$
That this is possible follows from
$M_0\in \Gamma^\perp$
(2.23), and from $M_+ \in \Gamma$ (and
hence $M_+ \in \Gamma^\perp$) which is
required by the Theorem in Section 3.
The conformal transformation (2.24) then reads
$$
\delta_f h=fh'+f'h+f''\,,
\qquad
\delta_f j_+=fj_+'+2f'j_+\,,
\qquad
\delta_f t=ft'+f'(t+[M_0,t])\,.
\eqno(4.3)$$
It will be useful to consider the subgroup of the
gauge group generated by $M_+$, which acts on $J(z)$
according to
$$
h \rightarrow h+2\alpha\,,
\qquad
j_+\rightarrow j_++\alpha'-\alpha h -\alpha^2\,,
\qquad
t\rightarrow e^{\alpha M_+} t e^{-\alpha M_+}\,.
\eqno(4.4)$$
A complete, polynomial gauge fixing of this gauge freedom
is obtained by restricting the current to the form
$$
I(z)=M_- + \omega(z) M_+ + u(z)\,,
\qquad\hbox{with}\quad
u(z)\in \Gamma^\perp\cap \S^\perp\,.
\eqno(4.5)$$
The general current (4.2) is transformed to this $M_+$-gauge
section by choosing the parameter $\alpha(z)$ to be
$\alpha(z)=-{1\over 2}h(z)$.
As a result, the differential polynomials given by
$$
u=e^{-{1\over 2}h M_+}t e^{{1\over 2}h M_+}\,,
\qquad
\omega = j_+ +{1\over 4} h^2 -{1\over 2}h'\,,
\eqno(4.6)$$
are invariant under the $M_+$-transformations (4.4),
and freely generate the ring
of the $M_+$-gauge invariant differential polynomials.
Since every $\Gamma$-gauge invariant differential polynomial
is necessarily $M_+$-gauge invariant (since $M_+ \in \Gamma$),
we see that every $P\in \R$ can be expressed in terms
of the $M_+$-gauge invariants, $u$ and $\omega$.
In more detail, we can write
$$
P=P(u,\omega)=\sum_i P_i(u,\omega) \,,
\eqno(4.7)$$
where the $P_i$ are uniquely determined differential polynomials
that are {\it homogeneous} of degree $i$ in their arguments
$u$, $\omega$.
This expansion is very convenient for investigating
the transformation properties of differential
polynomials under infinitesimal M\"obius transformations
($f'''=0$), because the expansion is covariant under
such transformations.
Indeed, from (4.3) and (4.6) we find
that $u$ and $\omega$ transform in a {\it linear,
homogeneous} way under M\"obius transformations:
$$
\delta_f u = f u' + f' (u+[M_0,u]) -{1\over 2} f''[M_+,u]\,,
\qquad
\delta_f \omega = f\omega' +2f'\omega\,.
\eqno(4.8)$$
(For completeness, we
note that under a general conformal transformation
$\delta_f \omega$ picks up also the usual $f'''$ term.)
Since the conformal transformation of a differential
polynomial is determined through the transformation of its
arguments, we obtain
$$
(\delta_f P)_i(u,\omega)=(\delta_f P_i)(u,\omega)\,.
\eqno(4.9)
$$
As a consequence, we find that
{\it a differential polynomial
$P(u,\omega)=\sum_i P_i(u,\omega)$ is
quasi-primary of scale dimension
$\Delta$ if and only if $P_i(u,\omega)$ is
quasi-primary of scale dimension $\Delta$ for all $i$.}
On account of this, we have the following general
idea for deriving restrictions on $\Gamma$
from requiring the existence of a quasi-primary basis in $\R$:
We should look at the linear, quadratic etc.~terms
in the expansion of the components of the DS gauge fixed current
$j_{\rm DS}(u,\omega)$ that generate $\R$,
and inspect the conditions
under which they can be expressed as differential
polynomials in homogeneous, quasi-primary
differential polynomials in $u$ and $\omega$,
since such differential polynomials enter the
expansion of the quasi-primary basis.

We shall see shortly how the above
idea works in the simplest linear case,
but before that we wish to mention
some further features of the gauge fixing with respect to $M_+$.
First, this partial gauge fixing is stable under
the subgroup of the gauge group generated by
$$
{\widetilde \Gamma}:=\Gamma\cap \S^\perp \,,
\eqno(4.10)$$
{\it i.e.}, by the subalgebra ${\widetilde\Gamma}\subset\Gamma$
defined by removing
$M_+$ from $\Gamma$ so that the rest is orthogonal to $M_-$.
The stability of the $M_+$-gauge section (4.5) under
the $\widetilde \Gamma$-gauge transformations can be seen
explicitly by
observing the $\widetilde \Gamma$-invariance
of the partial gauge
fixing condition,
$$
\la M_0, J(z)\ra=0\,,
\eqno(4.11)$$
that restricts the current to the form (4.5).
Second,  $\Gamma$ has the following semi-direct
sum structure:
$$
\Gamma={\rm span}\{M_+\}\oplus_s {\widetilde \Gamma}\,,
\qquad (\hbox{{\it i.e.}, $\,$}
[M_+,{\widetilde\Gamma}]\subset{\widetilde\Gamma})\,.
\eqno(4.12)$$
Accordingly, one can write the element
$g(z)=e^{\gamma(z)}$, $\gamma(z)\in \Gamma$,
of the gauge group in the product form
$$
g(z)=e^{\widetilde\gamma(z)} \cdot e^{\alpha(z) M_+}
\qquad\hbox{with}\qquad
\widetilde\gamma(z)\in {\widetilde\Gamma},
\eqno(4.13)
$$
and thereby fix the $M_+$-gauge-freedom first
in the way given above, and fix the
$\widetilde\Gamma$-gauge-freedom subsequently.
Having performed the first step, from now on we regard
$I(z)$ in (4.5) as our new variable,
whose components have the transformation rule (4.8) under
the M\"obius group and
upon which the further $\widetilde\Gamma$ gauge
fixings are to be performed.
The variables $u$, $\omega$ are simpler to deal with
than the original variables $t$, $j_+$, $h$,
since  the M\"obius group acts homogeneously
on the former (4.8) whereas it acts
inhomogeneously on the latter (4.3).

\medskip
\noindent
{\bf 4.2. Half-maximality of $\Gamma$ from the linear terms}
\medskip

Below we prove a proposition from which the dimensional estimate (4.1)
will follow as a corollary.
The proof will be based on a preliminary lemma,
which is an analogue of Lemma 1 of Section 3.

Let $q\in\{1, {3\over 2}, 2, {5\over 2}, \ldots \}$ be fixed and
(if exists) choose a nonzero element
$T_{-q}\in (\Gamma^\perp)_{-q}$.
As can be readily seen, we have
$$
({\rm ad\,}_{M_+})^{2q}(T_{-q})\neq 0\,.
\eqno(4.14)$$
Define ${\cal C}[T_{-q}]$ to be the following
subspace of the space of $M_+$-gauge fixed currents
given by (4.5):
$$
{\cal C}[T_{-q}]:=\{\, I\,\vert\,
I(z)=M_-+\sum_{i=0}^{i_{\rm max}} v_{i-q}(z)
({\rm ad\,}_{M_+})^i(T_{-q})\,\}\,,
\eqno(4.15)$$
where $i_{\rm max}$ is the largest natural number
for which  $({\rm ad\,}_{M_+})^{i_{\rm max}}(T_{-q})\neq 0$
(from (4.14) we have $i_{\rm max} \geq 2q$),
and the current components $v_{i-q}(z)$ are arbitrary.
In other words, the special configurations
${\cal C}[T_{-q}]$ are given by
the $M_+$-gauge-fixed current (4.5) where
all the components including
$\omega$ vanish, except for a single \lq $M_+$-string' of
$u$-fields, namely, the $v_{i-q}(z)$'s.
The point is that the subspace
${\cal C}[T_{-q}]$ is invariant under
the M\"obius transformations (4.8), which
act on $I(z)\in {\cal C}[T_{-q}]$ as
$$
\delta_f v_{i-q}=fv_{i-q}'+(1+i-q)f'v_{i-q}-{1\over 2}f'' v_{i-q-1}\,,
\quad \forall\,i=0,\ldots,i_{\rm max}\,,\quad (v_{-q-1}=0).
\eqno(4.16)$$
This means that, under this transformation,
the notion of quasi-primary differential
polynomials is well-defined even when restricted to
the subspace ${\cal C}[T_{-q}]$.

Let us set $b_q=0$ or $1 \over 2$ for $q$ integral or half-integral,
respectively. Then for any integer $0\leq k \leq q-b_q -1$,
the most general {\it linear} differential expression
of scale dimension $k+b_q+1$ that can be formed from
$I(z)\in {\cal C}[T_{-q}]$ is given by
$$
p_k = A_0 v_{k+b_q} + \sum_{i=1}^{q+k+b_q} A_i \pa^i v_{k+b_q-i}\,,
\eqno(4.17)$$
where the $A_i$ are arbitrary constants.
We then have the following auxiliary statement.

\medskip\noindent
{\bf Lemma.} {\it If the linear differential polynomial $p_k$ in
$(4.17)$ is quasi-primary on ${\cal C}[T_{-q}]$,
then $A_0 = 0$ for $0 \leq k \leq q-b_q-1$.}

\medskip
\noindent
{\it Proof.}
By computing the M\"obius transformation of $p_k$ in (4.17) through
(4.16), we obtain
$$
\delta_f p_k = fp'_k + (k+b_q+1)f'p_k + f''U(p_k)\,,
\eqno(4.18)$$
where
$$
U(p_k) = {1 \over 2} \; \sum_{i=1}^{q+k+b_q}
\bigl[ i(2k+2b_q+1-i)A_i - A_{i-1}\bigr]
\pa^{i-1}v_{k+b_q-i}\,.
\eqno(4.19)$$
For $p_k$ to be quasi-primary on ${\cal C}[T_{-q}]$
one must have $U(p_k)=0$
for any current $I(z)\in {\cal C}[T_{-q}]$.
The observation that the coefficient of $A_i$ in (4.19)
vanishes for
$i=2k+2b_q+1$ (which occurs for $0 \leq k
\leq q-b_q-1$) leads at once to $A_{2k+2b_q} = A_{2k+2b_q-1} =
\ldots = A_1 = A_0 = 0$.
{\it Q.E.D.}

\medskip
We now prove the main result of the section.

\medskip
\noindent
{\bf Proposition.}
{\it Suppose that there exist a DS gauge and a quasi-primary
basis in $\R$ (at the linear level).
Then for $q=1,{3 \over 2},2,{5 \over 2},\ldots$,
$\Gamma$ must satisfy the following relations:
$$
({\rm ad\,}_{M_+})^{q+k+b_q}\left((\Gamma^\perp)_{-q}\right)\subset
[M_-,\Gamma_{k+b_q+1}]\,,
\qquad \forall\,k=0,1,\ldots,q-b_q-1\  ,
\eqno(4.20)$$
where $b_q=0$ or $1 \over 2$ depending on whether $q$ is
integral or half-integral.}

\medskip
\noindent
{\it Proof.}
We can transform the $M_+$-gauge fixed current
$I(z)$ in
(4.5) to the fully DS gauge fixed current by a
${\widetilde \Gamma}$-gauge-transformation,
and the components of the resulting $j_{\rm DS}(u,\omega)\in V$
freely generate $\R$,
where $V$ is given in (2.11) which defines the DS gauge.
It follows from the differential
polynomial nature of the DS gauge fixing that,
when decomposed according to (4.7),
the linear terms of the components of $j_{\rm DS}(u,\omega)$,
defined by using some graded basis of $V$,
contain the corresponding components of $I(z)$.
We also know (cf.~Section 2) that the components of
$j_{\rm DS}(u,\omega)$ have definite scale dimensions.
{}From these facts it follows that if
(4.20) did not hold
for some $q$ and some $k$, then
we could find a component $P(u,\omega)\in \R$
of $j_{\rm DS}(u,\omega)$ whose linear term $P_1(u,\omega)$
reduces to an expression of the form (4.17) with $A_0\neq 0$
when restricted to a subspace ${\cal C}[T_{-q}]$
defined for a $T_{-q}$ with
$({\rm ad\,}_{M_+})^{q+k+b_q}(T_{-q})\notin
[M_-,\Gamma_{k+b_q+1}]$.
On the other hand, if there
exists a quasi-primary basis in $\R$, then
this $P(u,\omega)$ can be
expressed as a differential polynomial in
the basis,
and thus $P_1(u,\omega)$ must be a differential linear
combination of the quasi-primary linear terms of
the basis elements, that is,
$$
P_1(u,\omega) = Q_{k+b_q+1}(u,\omega) +
Q'_{k+b_q}(u,\omega) + Q''_{k+b_q -1}(u,\omega) + \ldots \ ,
\eqno(4.21) $$
where $Q_i(u,\omega)$ is a linear quasi-primary differential
polynomial of scale dimension $i$. Since $P_1(u,\omega)$ contains
the term $v_{k+b_q}$, $Q_{k+b_q+1}(u,\omega)$ must contain it as well.
Clearly, this
is a contradiction, because due to the Lemma
there is no such quasi-primary
differential polynomial of the form (4.7) whose
linear term contains a nonzero multiple of $v_{k+b_q}$
when restricted to ${\cal C}[T_{-q}]$.
We therefore conlude that (4.20) must hold.
{\it Q.E.D.}\footnote{$^3$}{
If $\g$ is known to be graded by the $sl(2)$ Casimir,
then for fixed $q$ the conditions in (4.20) for $k>0$ all follow
from that for $k=0$.}

\medskip
{}From the above Proposition, we easily obtain the
following dimensional bounds.

\smallskip
\noindent
{\bf Corollary.}
{\it For all $q \geq 1$ and $0 \leq k \leq q-b_q-1$, we have
$$
\d \g_q + \d
\g_{k+b_q+1} \geq \d \G_q.
\eqno(4.22)
$$
}
\noindent
{\it Proof.}
{}From (4.14) we obtain
$$
\d \Bigl[
  ({\rm ad\,}_{M_+})^{q+k+b_q}\left((\Gamma^\perp)_{-q}\right)
   \Bigr]
= \d (\Gamma^\perp)_{-q}=\d \G_{-q}-\d \Gamma_q =
\d \G_q-\d \Gamma_q.
\eqno(4.23)
$$
On the other hand, from (4.20) and the nondegeneracy condition
(2.16) we have
$$
\d \Bigl[
({\rm ad\,}_{M_+})^{q+k+b_q}\left((\Gamma^\perp)_{-q}\right)
\Bigr]
\leq \d [M_-,\Gamma_{k+b_q+1}] = \d \Gamma_{k+b_q+1}\,.
\eqno(4.24)
$$
Combining (4.23) and (4.24) we get (4.22).  The inequalities (4.1)
are recovered upon choosing $k=q-b_q-1$. {\it Q.E.D.}

\medskip
The relations (4.20) restrict the size
of $\Gamma$ considerably
as well as its position in $\G$ with respect to
the $sl(2)$ subalgebra $\S$.
To derive (4.20) we only used the requirement that
the {\it linear} terms of the
DS gauge fixed current should be expressible in terms
of the linear terms of a quasi-primary basis of $\R$.
It is plausible that by carrying on
the analysis to the quadratic and higher levels
one should obtain further restrictions on $\Gamma$
from the requirement of
the existence of a quasi-primary basis.
Unfortunately, it appears at the moment that such an analysis
does not yield a clearcut condition on $\Gamma$,
and for this reason this issue will not be pursued
further in this paper.

\vfill\eject

\centerline{\bf 5. Conformal spectrum and
decoupling in noncanonical DS reductions}
\medskip

Suppose that we construct a $\W$-algebra by using the DS mechanism
but {\it not} by a canonical DS reduction described in Section 2.2.
Suppose also that in the $\W$-algebra
{\it no negative conformal weight occurs} (in fact, so far we have
no example with a negative weight)
with respect to $L_{M_0}$.  We shall then show in
Section 5.1 that the $\Delta \geq {3\over2}$ part of the conformal
spectrum, given by the weights of the basis elements in the
$\W$-basis (or quasi-primary basis),
is {\it completely fixed} by the $sl(2)$ subalgebra $\S\subset\G$
associated to the reduction by the Theorem of Section 3.
In other words, the $\Delta \geq {3\over2}$ part of the conformal
spectrum
is the {\it same} as for the corresponding $\W_\S^\G$-algebra
obtained by canonical DS reduction.
In the subsequent Sections 5.2 and 5.3 we
show by examples that {\it there do exist noncanonical DS reductions},
where the resultant $\W$-algebras possess extra
`low-lying' weights $\Delta\in \{0, {1\over 2}, 1\}$
in addition to the canonical conformal spectrum of $\W_\S^\G$.
However, we find that these $\W$-algebras
are not essentially different from the
$\W_\S^\G$-algebras, since they decouple into the direct product
of a $\W_\S^\G$-subalgebra and a system of free fields.
It would be interesting to know
whether the decoupling mechanism we exhibit here in specific examples
works in other noncanonical DS reductions too.
As far as the decoupling of weight $1\over 2$ fields is concerned,
one may expect this to be a general phenomenon in
DS reductions by analogy
with the general decoupling theorem established
in the context of meromorphic conformal field theory by
Goddard and Schwimmer [32].

\medskip
\noindent
{\bf 5.1. Conformal spectrum from the $sl(2)$ embedding}
\medskip

Consider a $\W$-algebra resulting from DS reduction.
Let $V$ be the graded complementary space defining the DS gauge,
given in (2.11).
By the Theorem of Section 3, we can assume that
the grading is by the $sl(2)$ generator $M_0$.
We noted in Section 2 (see (2.25)) that the generators of
$\R$ provided by the components of $j_{\rm DS}(j)$ have
definite scale  dimensions obtained from the grades of the
basis of $V$ by a shift by $+1$.
This clearly implies that the spectrum of conformal weights
in any $\W$-basis (or quasi-primary basis) of $\R$, with respect to $L_{M_0}$,
is determined by the spectrum of $M_0$-grades in $V$ in the same way.

Let us now consider the case where
{\it no negative conformal weight occurs} in
our $\W$-algebra.
Note that we have the equality
$$
\d V_m = \d \G_m -\d\Gamma_{-m}-\d\Gamma_{m+1}\,,
\qquad \forall\, m\,,
\eqno(5.1)$$
on account of the decomposition (2.11)
and the nondegeneracy condition (2.16).
If we combine this equality with
the {\it nonnegativity} assumption,
$$
\d V_m =0
\qquad\hbox{for}\qquad m\leq -{3\over 2}\,,
\eqno(5.2)$$
then we get the formula
$$
\d V_m =\d \G_m -\d\G_{m+1}
\qquad\hbox{for}\qquad m \geq {1\over 2}\,.
\eqno(5.3)
$$
This tells us that the $\Delta \geq {3\over 2}$
sectors of the conformal weights
of the generators of our $\W$-algebra
are necessarily {\it the same} as for the
$\W_\S^\G$-algebra where $\S$ is the
$sl(2)$ containing $M_0$.
Thus the conformal spectrum can be different only
for the weight $0$ and ${1\over 2}$ sectors
(which do not exist in the canonical case), and
the weight $1$ sector.
By summing over all the grades in (5.1) and comparing it with
the corresponding sum taken for the canonical DS reduction, we derive
the formula
for the dimension of these sectors,
$$
\d V_{-1}+\d V_{-{1\over 2}} +
\bigl(\d V_0 -\d (V_{\rm c})_0\bigr)=
2(\d \Gamma_{\rm c} -\d\Gamma)\,.
\eqno(5.4)$$
Note that the dimension of $\Gamma_{\rm c}$ is the
maximal one allowed by the first-classness of
the constraints and the nondegeneracy condition (2.16),
and that $(V_{\rm c})_0$ is the space
of $sl(2)$ singlets in $\G$.
It is also useful to spell out from (5.1) the
dimensions of the extra sectors more explicitly,
$$
\eqalign{
\d V_0-\d (V_{\rm c})_0&=\d V_{-1}=\d\G_1-\d\Gamma_1-\d\Gamma_0\,,\cr
\d V_{-{1\over 2}}&=\d \G_{1\over 2} -2\d \Gamma_{1\over 2}\,.\cr}
\eqno(5.5)
$$
This means that we must have at least
as many conformal vectors as in the canonical case,
the number of extra conformal vectors
equals that of the conformal scalars,
and conformal spinors occur whenever
$\d\Gamma_{1\over 2}$ is smaller than in the canonical case.
We next present examples where such extra
low-lying fields indeed occur,
and shall see that, in those examples,
the $\W$-algebra decouples
into the direct product of a subalgebra isomorphic to $\W_\S^\G$ and
extra `free fields' of weight $0$, $1\over 2$ and $1$.

\medskip
\noindent
{\bf 5.2. Decoupling of weight $1\over 2$ fields}
\medskip

Consider a {\it half-integral} $sl(2)$
embedding $\S=\{M_-,M_0,M_+\}\subset \cal G$.
Recall that the canonical first class constraints
are defined by $\Gamma_{\rm c}$ in (2.28) and restrict
the current to the form given in (2.29).
In this section we are interested in
noncanonical DS reductions that are `marginal modifications'
of the canonical DS reduction obtained by removing some of the
canonical
constraints belonging to grade ${1\over 2}$ elements of
$\Gamma_{\rm c}$.
This means that our modified gauge subalgebra $\Gamma$ is of the type
$$
{\cal G}_{\geq 1} \subset \Gamma
 \subset \Gamma_{\rm c}\,,
\eqno(5.6)
$$
and the constraint surface $\K_\Gamma$
consists of currents of the form
$$
J(z)=M_-+j(z),\quad j(z)\in \Gamma^\perp,
\qquad\hbox{with}\qquad
\Gamma^\perp=(\Gamma^\perp)_{-{1\over 2}}+\G_{\geq 0}\,.
\eqno(5.7)$$
{}From the sufficient condition (2.21), the gauge group
admits the DS gauge fixing (with the grading defined by $M_0$)
and hence the corresponding ring $\R$ is freely generated.
It is also clear that $L_{M_0}\in \R$, but it is not
obvious whether there exists a $\W$-basis in $\R$.
However, one sees from (5.6) that if there is a $\W$-basis
in $\R$ then it must contain a subset of generators whose conformal
weights coincide with those of the $\W_\S^\G$-algebra and
$2(\d\Gamma_{\rm c} -\d \Gamma)$ additional conformal spinors
(see (5.5)).
In fact, below we shall exhibit two subrings,
$\R_{1\over 2}$ and ${\widehat \R}$, in $\R$, and the section
is devoted to proving the following statements:
\item{i)}
The subrings $\R_{1\over 2}$ and $\widehat \R$
are closed (in the usual local sense given in (2.9))
with respect to the induced Poisson bracket carried
by $\R$, and commute with each other under the Poisson bracket.
\item{ii)} The subring $\R_{1\over 2}\subset \R$ is freely
generated by a basis consisting of weight $1\over 2$
bosonic free fields.
\item{iii)}
The subring $\widehat \R$ is freely generated by a basis
subject to the $\W_\S^\G$-algebra under the Poisson bracket.
\item{iv)}
The union of the bases of $\R_{1\over 2}$ and $\widehat \R$
gives a basis of $\R$.
\item{v)}
The Virasoro generator
$L_{M_0}\in \R$ is the sum of the Virasoro
generators of the subrings $\R_{1\over 2}$ and $\widehat \R$,
$L_{M_0}={\cal L}_{1\over 2} +{\widehat {\cal L}}$.
\item{vi)}
The $\W$-basis of $\R$ is obtained from the
decoupled basis in iv) by replacing the Virasoro generator
${\widehat {\cal L}}$ of the $\W_\S^\G$-algebra carried by
$\widehat \R$ by $L_{M_0} \in \R$.
\smallskip

Let us begin by considering the subalgebra
$\widehat \Gamma\subset \G$ given by
$$
{\widehat \Gamma}:= {\widehat \Gamma}_{1\over 2} +\G_{\geq 1},
\eqno(5.8)$$
where  the subspace
${\widehat \Gamma}_{1\over 2}\subset \Gph$
is defined by
$$
[M_-,{\widehat \Gamma}_{1\over 2}]=(\Gamma^\perp)_{-{1\over 2}}.
\eqno(5.9)$$
One easily verifies the following
relations satisfied by the subalgebras introduced above:
$$
[{\widehat \Gamma},\Gamma]\subset\Gamma,
\qquad
[M_-,{\widehat \Gamma}]\subset \Gamma^\perp,
\qquad
{\widehat \Gamma}\subset \Gamma^\perp\,,
\eqno(5.10{\rm a})$$
$$
{\rm Ker\,}\left({\rm ad}_{M_-}\right) \cap {\widehat \Gamma}=\{0\}\,,
\eqno(5.10{\rm b})$$
$$
\Gamma^\perp =[M_-,{\widehat\Gamma}] +
{\rm Ker\,}\left({\rm ad}_{M_+}\right)\,.
\eqno(5.10{\rm c})$$
$$
\Gamma\subset\Gamma_{\rm c}\subset\widehat\Gamma\,.
\eqno(5.10{\rm d})$$
As we shall see shortly, the construction
will mainly depend on these relations.
Defining
$$
\phi_\alpha(z) :=\la \alpha,J(z)\ra -\la\alpha,M_-\ra\,,
\qquad \alpha\in\G,
\eqno(5.11)$$
we see that (5.10a) is equivalent to the equation,
$$
\{\phi_{\hat \gamma}(z),\phi_\gamma(w)\}\vert_{{\cal K}_\Gamma}=0\,,
\qquad \hat \gamma \in {\widehat \Gamma},\quad \gamma\in\Gamma\,.
\eqno(5.12)$$
This implies that the KM transformations generated by $\widehat\Gamma$,
$$
J\longrightarrow {\rm Ad\,}_{e^F} J:= e^F J e^{-F} +(e^F)' e^{-F},
\qquad  F(z)\in {\widehat \Gamma},
\eqno(5.13)$$
which contain the gauge transformations corresponding to
$F(z)\in \Gamma$,
are well-defined on the constraint surface ${\cal K}_\Gamma$
({\it i.e.}, preserve the form (5.7)).
Therefore we can define ${\widehat \R}\subset \R$ to be the
subring consisting of the ${\widehat \Gamma}$-invariant
(invariant under (5.13))
differential polynomials on ${\cal K}_\Gamma$.
It also follows from (5.12) that ${\widehat \R}$
is closed with respect to the
induced Poisson bracket carried by $\R$, {\it i.e.},
if $T, U\in {\widehat \R}$ then $P^i_{TU}$ in (2.9) belongs
to $\widehat \R$.
Furthermore, by writing $\widehat\Gamma$ in the form,
$$
{\widehat\Gamma}=\Gamma +\Sigma\,,
\qquad\hbox{with}\qquad \Sigma\cap\Gamma =\{0\}\,,
\eqno(5.14)$$
we obtain from (5.12) that the current components
$\phi_\sigma(z)$, $\sigma\in \Sigma$ are $\Gamma$-invariant
on $\K_\Gamma$ and hence belong to $\R$.
We define $\R_{1\over 2}$ to be the subring of $\R$ generated by these
current components. It is easy to see that the induced Poisson
bracket closes on $\R_{1\over 2}$ too in the usual local sense.
To finish the proof of statement i), we just
note that $\widehat\R$ and $\R_{1\over 2}$ commute with
each other under the Poisson bracket since
$\widehat\R$ consists of $\widehat\Gamma$-invariants, and
the current components $\phi_\sigma$, that generate
the differential ring $\R_{1\over 2}$ by definition,
generate infinitesimal $\widehat\Gamma$-transformations
through the Poisson bracket.

In order to establish ii), we make a concrete choice
for the space $\Sigma$ in (5.14) (the subring
$\R_{1\over 2}$ is easily seen to be independent
of the choice).
We do this by first choosing a subspace
$\Qph \subset\G_{1\over 2}$ on which the 2-form $\omega_{M_-}$
vanishes and for which $\Gph=\Pph +\Qph$,
with $\Pph$ appearing in the definition of $\Gamma_{\rm c}$ (2.28).
It follows that if we define the subspaces
${\cal P}, {\cal Q}\subset {\widehat \Gamma}_{1\over 2}$ by requiring
$$
\Pph=\Gamma_{1\over 2} +{\cal P},
\qquad
\hbox{and}\qquad
{\cal Q}:=\Qph \cap [M_-,\Gamma_{1\over 2}]^\perp\,,
\eqno(5.15{\rm a})$$
then we can take
$$
\Sigma:={\cal P}+{\cal Q}\,.
\eqno(5.15{\rm b})$$
These definitions guarantee that we can choose bases
$\{X_i\}\subset \cal P$, $\{Y_i\}\subset \cal Q$
so that we have
$$
\omega_{M_-}(X_i,Y_k)=\delta_{ik},
\qquad
\omega_{M_-}(X_i,X_k)=
\omega_{M_-}(Y_i,Y_k)=0.
\eqno(5.16)$$
The corresponding basis of $\R_{1\over 2}$ is given by
the current components
$$
p_i(z):= \phi_{X_i}(z)
\qquad\hbox{and}\qquad
q_i(z):= \phi_{Y_i}(z)\,,
\eqno(5.17)$$
whose Poisson brackets read
$$
\{p_i(z),q_k(w)\}\vert_{{\cal K}_\Gamma}=\delta_{ik}\delta(z-w),
\qquad
\{p_i(z),p_k(w)\}\vert_{{\cal K}_\Gamma}=
\{q_i(z),q_k(w)\}\vert_{{\cal K}_\Gamma}=0.
\eqno(5.18)$$
One readily checks that these elements of $\R$ are weight $1\over 2$
conformal
primary fields both with respect to $L_{M_0}$
and with respect to their own free field Virasoro density
${\cal L}_{1\over 2}$ given by
$$
{\cal L}_{1\over 2} :={1\over 2}\sum_i (p_i'q_i - p_i q_i').
\eqno(5.19)$$
Thus we have exhibited the basis of $\R_{1\over 2}$
claimed in statement ii).

To prove the main statement iii), observe that (5.10b) is the analogue
of the earlier nondegeneracy condition (2.16) and (5.10c)
is similar to decomposition (2.11) used to define a DS gauge.
This suggests that the subspace of currents ${\cal C}_{\rm hw}$
given by
$$
{\cal C}_{\rm hw}:=\{\,J\,\vert\,J(z)=M_-+j_{\rm hw}(z)\,,
\quad j_{\rm hw}(z)\in {\rm Ker\,}\left({\rm ad}_{M_+}\right)\,\}\,,
\eqno(5.20)$$
that defined the highest weight gauge for the canonical DS reduction,
is a global, polynomial section of the $\widehat\Gamma$-action
(5.13) on $\K_\Gamma$.
This follows if we show that the equation ({\it i.e.}, the
analogue of (2.13)),
$$
j\longrightarrow
{\rm Ad\,}_{e^F} j:= e^F(j+M_-) e^{-F} - M_- +(e^F)' e^{-F}=j_{\rm hw},
\eqno(5.21{\rm a})$$
with
$$
j(z) \in \Gamma^\perp,\quad F(z)\in {\widehat \Gamma},
\quad j_{\rm hw}(z)\in {\rm Ker\,}\left({\rm ad}_{M_+}\right)\,,
\eqno(5.21{\rm b})$$
has a unique, differential polynomial solution $F(z)=F(j(z))$.
Indeed, if this is so then the resultant $j_{\rm hw}(j(z))$
is also a differential polynomial in $j$ on account of
${\widehat\Gamma}\subset \G_{> 0}$, which implies that
${\rm Ad\,}_{e^F} j$ is a finite differential polynomial in $F$.
The construction then guarantees that the components of
$j_{\rm hw}(j)$
are $\widehat\Gamma$-invariants and freely generate $\widehat\R$,
in analogy with the way one constructs a basis of gauge invariants
through DS gauge fixing.
Although we could verify the unique, polynomial solubility
of (5.21) directly by a recursive procedure based on the grading
similarly as for DS gauge fixing [23],  it will be
advantageous to solve (5.21) by a two-step procedure
utilizing that ${\cal C}_{\rm hw}$ (5.20) is a gauge section
in the canonical case.
In the two-step procedure first we reduce the current
$j\in \Gamma^\perp$
to the canonical form $j_{\rm c}\in \Gamma_{\rm c}^\perp$ and then
employ
the usual DS procedure to the highest weight gauge fixing available
for the canonical DS reduction.
To implement this, we write
the current $j(z)\in \Gamma^\perp$ in the form
$$
j(z)=\sum_i p_i(z) [Y_i,M_-]+r(z),
\qquad r(z)\in \Gamma_{\rm c}^\perp,
\eqno(5.22)$$
where $\{Y_i\}\subset {\cal Q}$ is the basis introduced earlier,
$p_i$ is defined in (5.17), and we
used that $\Gamma^\perp =[M_-,{\cal Q}]+\Gamma_{\rm c}^\perp$.
Then we see that the first step is implemented by the KM transformation
$$
j\longrightarrow  {\rm Ad\,}_{e^{-p\cdot Y}}\, j:=j_{\rm c}(j).
\eqno(5.23{\rm a})$$
In the second step the resultant current $j_{\rm c}(j)
\in \Gamma_{\rm c}^\perp$
can be brought to the subspace ${\cal C}_{\rm hw}$ (5.20) by a
unique $\Gamma_{\rm c}$-transformation (which is a particular
$\widehat\Gamma$-transformation on account of (5.10d)) since
${\cal C}_{\rm hw}$ is known to represent a global gauge section
for the canonical DS reduction,
$$
j_{\rm c}\longrightarrow  {\rm Ad\,}_{e^{f_{\rm c}}}\,
j_{\rm c}:=j_{\rm hw}(j_{\rm c})\,,
\qquad\hbox{with}\qquad
f_{\rm c}\in \Gamma_{\rm c}\,.
\eqno(5.23{\rm b})$$
After this two step process, the group element
$e^F$ in (5.21) turns out to  be
$$
e^F=e^{f_{\rm c}} e^{-p\cdot Y},
\eqno(5.24)$$
where $f_{\rm c}=f_{\rm c}(j_{\rm c}(j))$ is a differential polynomial
in its argument.
This implies the unique solubility of (5.21) for $F$
since the group parameters
$F$ and $(f_{\rm c},p\cdot Y)$ are related to each other in a
 one-to-one,
polynomial manner on account of the grading.
{}From the unique, polynomial solubility of (5.21) we conclude
that the ring ${\widehat \R}$ is freely generated by the components of
$j_{\rm hw}(j)=j_{\rm hw}(j_{\rm c}(j))$,
whose number is $\d\G -2\d\Gamma_{\rm c}$
(notice that the function $j_{\rm hw}(j_{\rm c})$ appearing here is
the same as that occurring in the canonical case).

It is now not difficult to see that the Poisson bracket algebra
formed by the basis $\jhw(j)$ of the subring $\widehat {\R}$ is
isomorphic to the $\W_\S^\G$-algebra.  This follows from the
very fact that the components of $\jhw(j)$ are $\widehat {\Gamma}$--invariant
and
hence commute with the canonical constraints, {\it i.e.},
$\{ \phi_{\gamma_{\rm c}}(z) , \jhw(j(w)) \}\vert_{\Kg} = 0$
for $\forall\, \gamma_{\rm c} \in \gc$,
and from the fact that
on the subspace ${\cal C}_{\rm hw}$,
$\jhw(j(z))$ reduces to the highest weight gauge current $\jhw(z)$
defined in (5.20).
More explicitly, from the first fact we
observe that for $\jhw(j(z))$
the Dirac bracket defined for the set
of canonical second class constraints
specifying the constraint surface ${\cal C}_{\rm hw}\subset \K$
(5.20) is identical to the Poisson bracket,
$$
\{\jhw(j(z))\, ,\,  \jhw(j(w))\}
=\{\jhw(j(z))\, ,\, \jhw(j(w))\}^*
\qquad {\rm on} \quad {\cal C}_{\rm hw}.
\eqno(5.25)
$$
Then from the second fact we see that the r.h.s. of (5.25)
is equivalent to the Dirac bracket of the current components $\jhw(z)$
entering the definition (5.20),
$$
\{\jhw(j(z))\, ,\, \jhw(j(w))\}^* =
\{\jhw(z)\, ,\, \jhw(w)\}^*
\qquad {\rm on} \quad {\cal C}_{\rm hw}.
\eqno(5.26)
$$
As mentioned in Section 2,
the r.h.s.~of (5.26) forms the $\W_\S^\G$-algebra
after the change of the basis in which the $M_+$-component
of $\jhw(z)$ is replaced by
the Virasoro density $L_{M_0}(j_{\rm hw}(z))$.
(Here $j_{\rm hw}(z)$ simply means the current defined on the
subspace ${\cal C}_{\rm hw}$ (5.20) and is to be distinguished
from the function $j_{\rm hw}(j(z))$ defined on $\K_\Gamma$.)
By combining the last two equations, and noting that
$\{\jhw(j(z))\, ,\,  \jhw(j(w))\}$
is $\widehat\Gamma$-invariant and thus determined by
its restriction to the section ${\cal C}_{\rm hw}$,
we obtain the $\W_\S^\G$-basis of $\widehat\R$
required by statement iii) similarly as in the canonical case.
Namely, we modify the basis
provided by the components of $j_{\rm hw}(j)$
by replacing the $M_+$-component of $j_{\rm hw}(j)$ with the
Virasoro density ${\widehat {\cal L}}(j)\in \widehat \R$
given by
$$
{\widehat {\cal L}}(j):=L_{M_0}(j_{\rm hw}(j))=
{1\over 2}\la M_-+j_{\rm hw}(j),M_-+j_{\rm hw}(j)\ra\,,
\eqno(5.27)$$
where we observed that $\la M_0, j_{\rm hw}\ra=0$.

To demonstrate statement iv), we show
that $\{ p_i, q_i, \jhw(j) \}\subset \R$ is a basis of $\R$.
We do this by showing that any $\g$-invariant differential
polynomial $P(j)\in \R$ can be expressed as a differential
polynomial in this set.
For this purpose, it will be useful to
decompose the unique solution $F(j)\in \widehat\Gamma$ of (5.21)
into a sum according to (5.14),
$$
F =\epsilon +f\,,
\qquad\hbox{with}\qquad
\epsilon \in \Sigma,\quad  f\in \Gamma\,.
\eqno(5.28)$$
By substituting this into (5.21) and inspecting
the lowest grade part of this equation, we obtain
$$
\epsilon =\sum_i q_i X_i -\sum_i p_i Y_i\,,
\eqno(5.29)$$
where $p_i$, $q_i$ are the gauge invariant
components of $j$ defined by (5.17).
On account of these equations and $[\Sigma,\Gamma]\subset\Gamma$
which holds for grading reasons, we can write
$$
e^F=e^{\epsilon +f}= e^{q\cdot X -p\cdot Y} e^{\tilde f}\,,
\quad\hbox{with}\qquad
{\tilde f}\in \Gamma\,,
\eqno(5.30)$$
where ${\tilde f}={\tilde f}(\epsilon, f)$ is determined
by the Baker-Campbell-Hausdorff formula.
We then see by inverting (5.21) using (5.30) that $j\in \Gamma^\perp$
can be written in the form
$$
j=
{\rm Ad\,}_{e^{-{\tilde f}}}({\rm Ad\,}_{e^{p\cdot Y-q\cdot X}}
j_{\rm hw})
\,,
\eqno(5.31)$$
where ${\tilde f}$, $j_{\rm hw}$ are uniquely
determined differential polynomials in $j$.
If now $P(j)\in \R$ is an arbitrary $\Gamma$-invariant,
then we have
$$
P(j)=P(
{\rm Ad\,}_{e^{-{\tilde f}}}
({\rm Ad\,}_{e^{p\cdot Y-q\cdot X}}\,j_{\rm hw}))=
P({\rm Ad\,}_{e^{p\cdot Y-q\cdot X}}\, j_{\rm hw}).
\eqno(5.32)$$
This implies that the ring $\R$ is indeed generated
by the set $\{ p_i, q_i, \jhw(j) \}\subset \R$.
Of course, the number of the elements in the basis set is
$$
\dim \Sigma + \dim \G - 2\, \dim \gc
= {\rm dim\,}\G -2 \, {\rm dim\,}\Gamma,
\eqno(5.33)
$$
as required.
Having proved statement iv) for specific bases of the subrings
${\R}_{1\over 2}$, $\widehat {\R}$ by the above, the statement
obviously holds for any two such bases as well.

Concerning statement v), observe first the following chain
of the equalities:
$$
{\widehat {\cal L}}(j)=L_{M_0}(j_{\rm hw}(j_{\rm c}(j)))=
L_{M_0}(j_{\rm c}(j))=
{1\over 2}\la M_- +j_{\rm c}(j)\,,\,M_-+j_{\rm c}(j)\ra -
\la M_0, j'_c(j)\ra\,,
\eqno(5.34)$$
where all equalities are due
to definitions except the second one, which
is due to the $\Gamma_{\rm c}$-gauge-invariance of $L_{M_0}$ on the
constraint surface $\K_{\Gamma_{\rm c}}$ of the canonical DS reduction.
Then, by using (5.19) and (5.23a), it is a  matter of direct
verification to derive
$$
L_{M_0}(j):={1\over 2}\la M_-+j\,,\,M_-+j\ra -\la M_0\,,\,j'\ra
={\cal L}_{1\over 2}(j) +{\widehat {\cal L}}(j)\,,
\eqno(5.35)$$
as claimed in statement v).

Finally, since $L_{M_0}(j)\in \R$ is linear in
${\widehat {\cal L}}\in \widehat \R$, statement vi) is
now obvious from the above.

In summary, in this section we have shown that the reduction belonging
to $\Gamma$ (5.6)  leads to a ${\cal W}$-algebra
that is isomorphic
to the direct product of the ${\cal W}_\S^\G$-algebra with a system
of weight $1\over 2$
bosonic free fields. The number of the $(p,q)$ pairs is
${1\over 2}{\rm dim\,}(\G_{1\over 2})$ in the exreme case
when $\Gamma={\cal G}_{\geq 1}$, and $0$
in the other extreme case $\Gamma=\Gamma_{\rm c}$.
Obviously, the systems obtained
by adding such free fields to the $\W_\S^\G$-algebra
cannot be considered genuinely new ${\cal W}$-algebras.
The above construction whereby we have seen
the decoupling mainly depended on the properties
collected under (5.10), but at some points also
on the specific grading structure of our example.
In particular, the fact that
$\Gamma$ in (5.6) differs from $\gc$ only by elements in
${\cal G}_{1 \over 2}$ is
a sufficient condition for the construcion to work in general.
Nevertheless, this
construction could perhaps serve as a `prototype'
in a more general study of noncanonical
DS reductions of $\Gamma\subset\Gamma_{\rm c}$ type (we have no
other kind of noncanonical example).
Although the range of
validity of this type of construction is not clear,
it is certainly not restricted to the above family of examples,
as is illustrated by a new example in the next section.

\medskip
\noindent
{\bf 5.3. Decoupling of weight $(0,1)$ fields}
\medskip
The modifications of the canonical DS reductions described in
the previous section were obtained by removing some of
the canonical constraints belonging to lowest grade
elements of $\Gamma_{\rm c}$ in the case of a half-integral $sl(2)$ embedding.
In the case of an integral $sl(2)$ embedding the same idea
cannot be applied {\it in general}, since the DS gauge fixing
would not be applicable for the modified system of constraints.
There are however particular cases where the idea works,
and we here present a simple example based on the
$sl(2)$ subalgebra of the Lie algebra $B_2$ belonging to a
short root.
We shall see that the modified reduction
leads to a $\W$-algebra that
decouples into the direct product of the corresponding
$\W_\S^\G$-algebra and a $(p,q)$ pair of free fields with
conformal weights $(0,1)$, quite analogously to what we have seen in
Section 5.2.

The root diagram of the Lie algebra $B_2$ consists of the vectors
$$
\pm e_1, \quad
\pm e_2,\quad
\pm (e_1\pm e_2)\,.
\eqno(5.36)$$
The algebra is spanned by the step operators and
the Cartan elements,
$$
E_{\pm e_1},
\quad
E_{\pm e_2},
\quad
E_{\pm (e_1\pm e_2)}\,,
\quad
H_{e_1},
\quad
H_{e_2}\,,
\eqno(5.37)$$
which we normalize by $[H_{e_i},E_{e_i}]=E_{e_i}$.
We consider the $sl(2)$ subalgebra belonging to the short
root $e_1$,
$$
M_\pm:=E_{\pm e_1},
\quad
M_0 :=H_{e_1}\,.
\eqno(5.38)$$
For the corresponding canonical DS reduction we have
$$
\Gamma_{\rm c}={\rm span\,}\{ E_{e_1+e_2}, E_{e_1}, E_{e_1-e_2}\,\}\,,
\eqno(5.39)$$
and
$$
{\rm Ker\,}\left({\rm ad\,}_{M_+}\right)
={\rm span\,}\{ E_{e_1+e_2}, E_{e_1}, E_{e_1-e_2}, H_{e_2}\,\}\,.
\eqno(5.40)$$
The adjoint representation decomposes under the $sl(2)$
according to $10=3\times 3 + 1$.
The first class constraints of the
modified reduction are determined by the pair
$(\Gamma,M)$ where we define $M:=M_-$ and
$$
\Gamma:={\rm span\,}\{ E_{e_1+e_2}, E_{e_1}\}\,.
\eqno(5.41)$$
One can directly check that the DS gauge fixing
is applicable in this case.
To see the structure of the reduced system we proceed
analogously as in Section 5.2.
We define
$$
{\widehat \Gamma}:= \Gamma + \Sigma\,,
\quad\hbox{\rm with}\quad
\Sigma:={\rm span\,}\{ E_{e_1-e_2}, E_{e_2}\}\,,
\eqno(5.42)$$
and then the analogues of the relations in (5.10)
are satisfied.
By using these relations we can verify also in this case
that the Poisson bracket algebra carried by the
ring, $\R$, of $\Gamma$-invariant
differential polynomials decouples into the direct product
of the $\W_\S^\G$-subalgebra carried
by the subring, ${\widehat \R}$, of ${\widehat\Gamma}$-invariants,
and the $\Gamma$-invariant currrent components
$$
p(z):={1\over {\sqrt2} }\la E_{e_1-e_2},J(z)\ra
\qquad\hbox{and}\qquad
q(z):={1\over {\sqrt2}}\la E_{e_2},J(z)\ra
\eqno(5.43)$$
generating another subring, $\R_{(0,1)}$.
On the constraint surface defined by $\Gamma$, these current
components satisfy the analogue of (5.18)
(since $\la M_- , [ E_{e_1-e_2} , E_{e_2}]\ra = 2$ in our convention).
The notation $\R_{(0,1)}$
reflects the fact that in this case $p$ is a conformal scalar
and $q$ is a conformal
vector with respect to $L_{M_0}\in \R$.
These conformal weights are assigned to the pair $(p,q)$
by the  quadratic Virasoro density
given by
$$
{\cal L}_{(0,1)}:=p'q\,,
\eqno(5.44)$$
and $L_{M_0}\in \R$ decomposes into the sum
of this Virasoro density and that of the $\W_\S^\G$-subalgebra,
similarly
as for the weight ${1\over 2}$ fields in Section 5.2.

\vskip 0.4truecm

\bigskip
\centerline{\bf 6. Discussion}
\medskip

The purpose of the present paper was to gain a better understanding
of the DS reduction approach to classical $\W$-algebras in general,
and in particular to investigate the  completeness of
the $\W_\S^\G$-algebras in the set of $\W$-algebras that may
be obtained from DS reductions.
On the basis of the definition given in Section 2,
we proved in Section 3 that all DS reductions can be determined by
triples
of the form $(\Gamma,M=M_-,H=M_0)$,
where $\S=\{M_-,M_0,M_+\}$ is an $sl(2)$ subalgebra of the underlying
simple Lie algebra $\G$ and $M_+\in \Gamma$.
This way we reduced the problem of listing all DS reductions
to the problem of finding all possible $\Gamma$'s for given
$sl(2)$ embeddings (whose classification is known).
Then we went on to exhibit restrictions on the allowed $\Gamma$'s
in Section 4.
The basic idea there was
that by inspecting the expansion of
the DS gauge fixed current and requiring term by term that it be
compatible with the existence of a quasi-primary basis in $\R$
one obtains conditions on $\Gamma$.
We completed the analysis only at the linear
level, but it should be possible to
pin down $\Gamma$ more closely by analysing the quadratic and
higher order terms of the expansion.
We also wish to emphasize at this point that the linear conditions
on $\Gamma$ given by the Proposition in Section 4.2 are to be
combined with the requirements imposed on $\Gamma$ by the
first classness of the constraints together with the
severe restriction for
the existence of a DS gauge.
All in all, we think $\Gamma$ is already very much constrained
by these conditions, but further study would be needed to have
the allowed $\Gamma$'s under complete control, ideally by deriving
their list.

We left the previous train of thoughts in Section 5 to some extent.
We there first showed that
if the conformal weight spectrum resulting from a DS reduction
is nonnegative then its
$\Delta \geq {3\over2}$ subsector is necessarily the same
as that of the corresponding $\W_\S^\G$-algebra.
We then found examples of new, noncanonical DS reductions,
which in principle yield new $\W$-algebras for which extra low-lying
weights $\Delta\in \{0, {1\over 2}, 1\}$ do occur.
However, in the examples we also found a mechanism
whereby the resulting $\cal W$-algebras
were identified as direct products of $\W_\S^\G$-algebras and systems
of free fields, {\it i.e.}, they
turned out to be not essentially new.

Our theorem on an $sl(2)$ embedding being associated
to every DS reduction and our result on the conformal weights
are consistent with the more abstract
results in [31] where an embedding of the
M\"obius $sl(2)$ into a finite Lie algebra
was associated to every classical $\W$-algebra
with positive, half-integral conformal spectrum
by using completely different methods.
More precisely, in [31] the classical $\W$-algebra
was viewed as the limit of a quantum one.  This led
to some unnecessarily restrictive
assumptions, which we removed in a recent preprint [33].
But, even taking this into account,
the assumptions in [31] and in the present work
are different.
For instance, refs.~[31,33] exclude
conformal scalars (and spinors),
which are some of the free fields that occur in our examples.
It is known that our `DS $sl(2)$ embedding'
and the `M\"obius $sl(2)$ embedding'
of [31] are isomorphic for
the {\it canonical} DS reductions
[31,33], but
it is not clear that they are isomorphic for {\it all}
possible {\it non}canonical DS
reductions which are the cases in which we are interested here.
The exact relationship between the results in [31,33]
and the present paper will be
clear when a more complete classification of
$\W$-algebras
and DS reductions becomes
available.

Pending such a complete classification, the results
derived in this paper give a strong support to
the {\it conjecture} that
{\it the set of $\W$-algebras with nonnegative
spectra $\Delta\geq 0$ that may be obtained from DS
reductions is exhausted by the
$\W_\S^\G$-algebras and decoupled
systems consisting of $\W_\S^\G$-algebras
and systems of free fields.}
On the basis of the results in [31,33],
it is also natural
to ask whether the $\W_\S^\G$-algebras
are exhaustive even outside the DS approach.

We wish close this paper by mentioning some other
open questions related to DS reductions, and to the above
conjecture.
First, let us recall that
the definition of the classical $\W$-algebra (and that
of the DS reduction) assumes a
preferred Virasoro density.
In view of the notion
of isomorphism between classical $\W$-algebras,
we are naturally led to the following basic questions:

\item{1.}
Are there nontrivial possibilities for finding two
$\W$-bases, $\{W_a\}$ and $\{\tilde W_a\}$, both freely generating an
invariant ring $\R$, such that the weights $\Delta_a$,
$\tilde\Delta_a$, and the centres $c$, $\tilde c$, relative to $W_1$
and $\tilde W_1$ respectively, are not identical?

\item{2.}
Are there `accidental isomorphisms' between $\W_\S^\G$-algebras
belonging to group theoretically {\it inequivalent} $sl(2)$ embeddings?

\noindent
We note here that the conformal structure
is not unique in a rather trivial way in the cases where the
$\W$-basis contains a $(p,q)$ pair,
$\{ p(z), q(w)\}=\delta(z-w)$, decoupled from the rest, since one can
assign conformal weights $(h, 1-h)$ to the pair with any $h$ by
building an appropriate
quadratic Virasoro density out of $p$, $q$.

Second, we used in Sections 3 and 4 the notion of a quasi-primary
basis, which is {\it in principle weaker} than the
notion of $\W$-basis, and derived conditions from its existence
by looking only at
the {\it linear} part of the ring $\R$.
Then in Section 5 we derived results from the
assumption that the conformal spectrum
is nonnegative,
and constructed noncanonical DS reductions which were
all found to lead to decoupled systems containing a
$\W_\S^\G$-subalgebra, but we made no attempt
to establish these results more generally.
In fact, these questions are open:

\item{3.}
What is the full set of conditions implied by the
existence of a $\W$-basis in the invariant ring $\R$?
\item{4.}
Does every  $\W$-algebra obtained from a noncanonical DS reduction
contain a $\W_\S^\G$-subalgebra? If it is so, is such an
algebra always `completely reducible'?
\item{5.}
Do DS reductions exist with {\it negative} conformal
weights occurring in the $\W$-basis of $\R$
with respect to $L_{M_0}$?

Third, the existence of a DS gauge is the only
{\it sufficient} condition we are aware of whereby
one can guarantee the invariant ring $\R$ to be freely generated.
In fact, we have no nontrivial
example for
$\R$ being freely generated without
the applicability of DS gauge
fixing.
Hence we should ask the following question:

\item{6.}
Are there other sufficient conditions than
the existence of a DS gauge
for ensuring that the invariant ring $\R$ is freely generated?

We also wish to note that in most
KM reductions by first class constraints
$\R$ may not be freely generated, simply by a
genericity argument.
(We explicitly show the non-existence of a free generating set
for the examples in the Appendix.)
Moreover, if the reduction is by conformally
invariant first class constraints then $\R$ may
in general be generated by invariants that include
a Virasoro density and
are subject to differential polynomial relations.
Thus there is a large set of extended conformal algebras built
on generating fields obeying differential--algebraic constraints
that one may derive from KM reduction, and
it is an open question whether one can or cannot make sense
of quantum versions of such algebras.

\vskip 0.7cm
\noindent
{\bf Acknowledgements.}
L. F. has been supported by the the Alexander von Humboldt-Stiftung.
He also wishes to thank W. Eholzer, A. Honecker,
R. H\"ubel, J. M. Figueroa-O'Farrill and W. Nahm
for useful comments.

\vskip 0.7cm
\noindent
{\bf Note added.}
After the first version of this paper was submitted
there appeared a preprint [35] containing
a decoupling algorithm and ref.~[36] dealing with an
 example of classical coset construction where the
analogue of the ring $\R$ is {\it infinitely} generated.
With regard to an aspect of the $sl(2)$ structure,
we also wish to mention refs.~[37,38].

\vfill\eject
\centerline{\bf Appendix: $W_{2n}^2$ examples of nonfreely
generated rings}
\medskip

In this appendix we consider
the $W_{2n}^2$-algebras of [29] and show by inspection
 that the corresponding ring $\R$ is
{\it not freely generated}.
These examples illustrate the difficulties one has to face in general
if one wants to describe the structure of the invariant ring for KM
reductions for which DS gauges do not exist.
(These difficulties appear similar to the ones encountered
in the general case of the GKO coset construction [3]).

As discussed in [30], the $W_{2n}^2$-algebras can be obtained
by reducing the KM algebra of $\G=sl(2n)$ by first class
constraints of type (2.5) with $(\Gamma,M)$ being the following.
Consider the $sl(2)$ subalgebra $\S=\{M_-,M_0,M_+\}\subset
 sl(2n)$ under which the $2n$ dimensional representation
decomposes into $2n=n+n$, and note that
the singlets of $\S$ in the adjoint of $sl(2n)$ form another
 $sl(2)$ subalgebra $\sigma=\{m_-,m_0,m_+\}\subset sl(2n)$.
The gauge algebra $\Gamma$ of the required first class constraints
 is given by the semidirect sum
$$
\Gamma={\rm span}\{m_+\}\oplus_s \Gamma_{\rm c}\,,\qquad
 ([m_+,\Gamma_{\rm c}]\subset \Gamma_{\rm c})\,,
\eqno({\rm A}.1)$$
where $\Gamma_{\rm c}\subset sl(2n)$ is the canonical
subalgebra (2.28) belonging
 to $\S$, and $M=M_-$.
The DS gauge fixing is not applicable to these first class constraints
 since,  on account of $[M_-,m_+]=0$, the nondegeneracy
condition (2.16) is not satisfied.
More precisely, the
$\Gamma_{\rm c}\subset \Gamma$ part of the gauge freedom can still be
fixed in
the usual differential polynomial way, and after doing this [30]
the independent components of the partially gauge fixed current may
 be displayed as
$$
\pmatrix{e&b\cr
         0&-e\cr}\,,
\qquad
\pmatrix{S_i+E_i&B_i\cr
         C_i&S_i-E_i\cr}\,,
\quad i=1,\ldots, (n-1)\,.
\eqno({\rm A}.2)$$
Except $S_1$ that enters  the Virasoro $L:=L_{M_0}=e^2 +S_1$ linearly,
all these fields are primary; $e$, $b$ have conformal weight $1$,
the fields with index $i$ have conformal weight $(i+1)$.
These components are differential polynomials in the
 original first class
constrained current, since they were obtained
by applying the standard DS gauge fixing to the $\Gamma_{\rm c}$ gauge
freedom.
It also follows [30] that the components
$$
e\,,\qquad S_i\,,\qquad C_i\,,
\eqno({\rm A}.3)$$
are invariant under the residual gauge transformations generated by
$m_+\in \Gamma$,
while the rest transforms according to
$$\eqalign{
E_i&\longrightarrow E_i +\alpha C_i\,,\cr
B_i&\longrightarrow B_i -2\alpha E_i -\alpha^2 C_i\,,\cr
b&\longrightarrow b-2\alpha e +\alpha'\,.\cr}
\eqno({\rm A}.4)$$

Thus the problem of finding a generating set for $\R$ is reduced
to the problem of
finding a generating set for the differential polynomial
invariants  in the components in ({\rm A}.2)
under the very simple gauge transformation rule ({\rm A}.4).
We below investigate this problem by using the following
observations.
First, notice that $\R$ is {\it graded by scale dimension},
{\it i.e.},
the homogeneous pieces with respect to scale dimension
belong to $\R$ separately for any element of $\R$.
One sees this for example from the fact that the gauge
transformation ({\rm A}.4) preserves scale dimension for
scalar $\alpha$.
(One could identify $\R$ as a certain factor-ring, and see
its being graded by scale dimension from that too.)
Thus it is natural to look for
a {\it homogeneous} generating set in $\R$,
{\it i.e.},
one consisting of elements having definite scale dimensions.
Second, because
the basic ingredients in ({\rm A}.2) from which the elements of $\R$
are
constructed have positive scale dimensions,
one sees that $\R$ is {\it positively graded},
and the subspaces of $\R$ with fixed scale dimension
have finite dimension.

This implies that if we want to select a
homogeneous generating set,
we can proceed by starting from the elements of
lowest scale dimension in $\R$
and include at each scale dimension
a minimal set of elements in the generating set
in such a way that the elements of $\R$ up to that scale dimension
are differential polynomials in these elements and the elements of
lower scale dimension.
We can implement this procedure by inspection up to some finite
scale dimension.
On the other hand, if there is  a basis (free generating set)
in $\R$ then the number of basis elements cannot be greater than
the number of degrees of freedom in the reduced system (obtained by
simple counting).
Hence we can conclude the {\it nonexistence of a homogeneous basis}
in $\R$
either $(a)$ if we have collected as many generators as the
number of reduced degrees of freedom and then exhibit an element of
$\R$ that cannot be expressed as a differential polynomial in these
generators, or $(b)$ if we find relations between the selected
generators  after having completed
the selection up to a given scale dimension.
By using this reasoning,
we shall find that in the cases we consider the ring
$\R$ {\it does not admit a homogeneous basis}.
As a consequence, it does not admit a $\W$-basis, since that
would be a particular homogeneous basis.
 We think the nonexistence of a {\it homogeneous} free
generating set implies that $\R$
does not admit {\it any} free generating set,
but this will not be shown here.
(We should note that the nature of the generating set of $\R$ has not
been investigated  so far, although the analogues of
the above first class constraints
and a differential rational gauge fixing
procedure were given in [30] for $W_k^l$ in general.)
We first consider the simplest case $n=2$, {\it i.e}, $W_4^2$.

\smallskip
\noindent
{\it i) The case $W_4^2$}
\smallskip

In this case the reduced system contains 5 degrees of freedom
since the number of fields in ({\rm A}.2) is now 6 and we have the one
parameter
gauge freedom ({\rm A}.4).
It is clear that the 3 gauge invariant components
$e$, $S$, $C$ must be included
in the generating set of $\R$ we are looking for.
(We suppress  the index $i$ in ({\rm A}.2-4), which in the present
 case
takes only the value $1$.)
The next simplest gauge invariants will be obtained by means of the
rational gauge fixing
$$
E\longrightarrow E +\alpha C =0
\qquad \Longrightarrow \qquad \alpha=-{E\over C}\,.
\eqno({\rm A}.5)$$
By plugging back this value of the gauge parameter into ({\rm A}.4),
we obtain
the following $2$ differential rational gauge invariants:
$$\eqalign{
B &\longrightarrow ({{E^2 + BC}})/C := R_1\,,\cr
b &\longrightarrow (bC^2 +2e EC +(EC'-E'C))/C^2 :=R_2\,.\cr}
\eqno({\rm A}.6)$$
By a similar argument used for a DS gauge fixed current (see
(2.14)),
it is easy to see that it is possible to express all
differential rational invariants, and thus also
the elements of $\R$ since polynomials are special rationals,
as differential rational functions in the components of the gauge
fixed current resulting from the rational gauge fixing.
In particular,
observing that the denominators in (A.6) are invariants themselves,
we obtain the elements of $\R$ given by the numerators
of $R_1$, $R_2$,
$$
X:={E^2 + BC}\,,
\qquad
P:={bC^2 +2e EC +(EC'-E'C)}\,.
\eqno({\rm A}.7)$$
By inspection, it is not hard to see that
there are in fact no simpler ({\it i.e.}, ones with lower scale
dimension)
elements of $\R$ in terms of which we could express
$X$ and $P$, which contain $B$ and $b$, respectively.
{}From
 this and the fact that the number of degrees of freedom is $5$,
we conclude that either the set $\{\, e, L, C, X, P\,\}$
is a  homogeneous basis for $\R$, or otherwise
there is no such free generating set in $\R$.
(If this was a free generating set then it
was also a $\W$-basis.
For this reason we exchanged the generator $S$ for the Virasoro
$L=e^2 +S$, which is obviously allowed.)

Let us next observe that the following combination of the rational
invariants
$$\eqalign{
K:=C^3 R_2^2 -(C')^2 R_1 &= b^2 C^3 + 4 e b E C^2 + 4 e^2 E^2 C
+2bC(EC'-E'C) \cr
&+4e (E^2 C' -EE'C)-B(C')^2\cr}
\eqno({\rm A}.8)$$
is a differential {\it polynomial} belonging to $\R$.
Since only $P$ contains $b$ in the combination $b C^2$,
we see that $K\in \R$
cannot be expressed as a differential polynomial
in the set $\{\, e, L, C, X, P\,\}$.
Therefore  $\R$ does not admit a (homogeneous) free generating set.
Consistently, if we now say that the generating set of $\R$ will be
$\{ e, L, C, X, P, K,\ldots \,\}$
then we receive $1$ differential polynomial
relation between the first $6$ generators,
$$
P^2 -KC -(C')^2 X =0\,.
\eqno({\rm A}.9)$$
Observe also that this generating set will not consist of
a Virasoro and primary fields, since $K$ in (A.8) is not a primary
 field.
Moreover, it should be stressed that we have no argument even for the
existence of a {\it finite} generating
set in $\R$!
We are not sure if a finite generating set exists in this case or not,
but it is conceivable for example that if we consider the elements of
$\R$ only up to some
finite scale dimension, then we find a generating set for that part
consisting of $g$ elements with $r$ relations in such a way that
$g-r$ is always $5$, but both $g$  and $r$ tend to $\infty$ as we
increase the scale dimension.

One can actually see already from the $W_4^2$ case that $\R$ is never
freely generated for $W_{2n}^2$ because a similar argument may be
applied
to those cases, too.
But it is worth also having a  closer look at the
next case, where this can be seen even without considering such a
 `tricky object'
as $K$ above (the `trick' there being the cancellation of the terms
proportional
with ${1/C}$, that are present before the substraction in ({\rm A}.8)).

\smallskip
\noindent
{\it ii) The case $W_6^2$}
\smallskip

The number of reduced degrees of freedom given by simple
counting is in this case $9$.
We have  now  5 linear invariants in ({\rm A}.3).
By  using rational gauge fixing or just looking at the transformation
rule of the capital letters in ({\rm A}.4),
we obtain the following $4$ quadratic invariants:
$$
\eqalign{
X_1:&=E_1^2 +B_1 C_1\,,\cr
X_2:&=E_2^2 +B_2 C_2\,,\cr}
\qquad
\eqalign{
X_{12}:&=2E_1 E_2 +B_1 C_2 +C_1 B_2\,,\cr
Y:&=C_1 E_2 -C_2 E_1\,.\cr}
\eqno({\rm A}.10)$$
It is clear that we have to include all the above linear and quadratic
invariants in
 the generating set of $\R$.
This already implies that $\R$ cannot be freely generated since we
 can verify the relation
$$
Y^2 -C_1^2 X_2 -C_2^2 X_1 +C_1 C_2 X_{12}=0\,.
\eqno({\rm A}.11)$$
Let us nevertheless continue the selection of the generating set a bit further.
So far we have $9$ generators and $1$ relation
and none of the generators we already have contains the component $b$.
The simplest
invariants involving $b$  are the following $3$ analogues of
$P$ in ({\rm A}.7),
$$\eqalign{
P_1:&=bC_1^2 +2eC_1E_1 +(C_1'E_1-C_1E_1')\,,\cr
P_2:&=2bC_1C_2+2e(E_1C_2+E_2C_1)+
(C_1'E_2+C_2'E_1-C_1E_2'-C_2E_1')\,,\cr
P_3:&=bC_2^2 +2eC_2E_2 +(C_2'E_2-C_2E_2')\,.\cr}
\eqno({\rm A}.12)$$
It is easy to see  that we also have to include these 3 invariants
in the generating set of $\R$
we are looking for.
Together with these $3$ generating elements  we receive
also $2$ new relations:
$$\eqalign{
C_1C_2P_2-C_2^2 P_1 -C_1^2 P_3 +(C_1 C_2'-C_1'C_2)Y&=0\,,\cr
C_1^2 P_3 -C_2^2 P_1 +(C_1C_2)Y'-(C_1C_2)'Y-2eC_1C_2Y&=0\,.\cr}
\eqno({\rm A}.13)$$
Thus the counting of degrees of freedom, $9=12-3$,
is `correct' at this stage,  though the set
$$
\{ e, L, C_1, S_2, C_2, X_1, X_{12}, Y, X_2, P_1, P_2, P_3\}\,,
\eqno({\rm A}.14)$$
where we exchanged $S_1$ for $L$, is not a generating set of $\R$.
Indeed,
in addition to invariants like $K$ in ({\rm A}.8),  one may check that
for example the following elements of $\R$ cannot
be expressed as differential polynomials in this set,
$$\eqalign{
T_{1}:&=C_2[bC_1^2+(C_1'E_1-C_1E_1')]^2
         +4eC_1^2E_2[bC_1^2+(C_1'E_1-C_1E_1')] -4e^2C_1^4B_2\,,\cr
T_{2}:&=C_1[bC_2^2+(C_2'E_2-C_2E_2')]^2
         +4eC_2^2E_1[bC_2^2+(C_2'E_2-C_2E_2')] -4e^2C_2^4B_1\,.\cr}
\eqno({\rm A}.15)$$
By adding these $2$ invariants to the generating set, we
also receive $2$
new relations, and it is not clear to us if the procedure would
terminate at a certain higher scale dimension or not.
The only firm concusion we can draw from the above is that $\R$ is
not freely generated for $W_6^2$ and that the generating set
(whatever it is) is pretty complicated.

\smallskip
\noindent
{\it iii) Further remarks}
\smallskip
Some further remarks are now in order.
First, the analysis given above clearly implies that $\R$ is
not freely generated also for any $W_{2n}^2$.
For example, one
can see this for any $n\geq 3$ simply by looking only at the
corresponding
linear and quadratic invariants.
The analogous statement is likely to be true for any $W_k^l$ ($1<l<k$),
except the cases $W_{2n+1}^2$ which coincide with particular
$\W_\S^\G$-algebras for $\G=sl(2n+1)$.
More generally, we may expect the structure of the
invariant ring $\R$ to
be similarly complicated for a generic KM reduction by first class
constraints.
(For further study of the structure of these complicated rings,
one can find references
on the mathematical literature on differential rings in [34].)

Finally, we wish to note that the above considered
reduction of the KM algebra by first class constraints
can be naturally  reinterpreted as the following two-step
reduction procedure [30].
The first step consists in reducing the KM algebra to the
$\W_\S^\G$-algebra
with $\G=sl(2n)$ and $\S$ given at the beginning of the appendix.
The second step consists in further reducing the $\W_\S^\G$-algebra
by using its sub-KM algebra given by the singlets (see also the remark
at the end of Section 2.2.).
This sub-KM algebra is now just the $sl(2)$ KM algebra
of the components belonging to
$\sigma =\{m_-,m_0,m_+\}\subset {\rm Ker}({\rm  ad}_{M_+})$.
The `secondary reduction' of the $\W_\S^\G$-algebra
({\it i.e.}, the second step of the KM reduction)
has been defined by putting the $m_+$-component ---
the lower-left entry of the first matrix in eq. ({\rm A}.2) ---
to the degenerate $0$ value.
The reader might wonder what happens if one puts that component to $1$,
rather than $0$, which means that one would perform
a DS reduction on the singlet sub-KM algebra
as the secondary reduction of the $\W_\S^\G$-algebra.
In fact, one can verify that this
`DS reduction after DS reduction' gives
no new kind of algebra;
it leads just to the $\W_{\tilde {\S}}^\G$-algebra where
${\tilde \S}=\S +\sigma$,
with the sum applied to the $sl(2)$ generators.
Clearly, this has a natural generalization, that is, DS reductions
of $\W_\S^\G$-algebras to other canonical algebras
as far as there is a semisimple part in the singlet
KM to perform a secondary
DS reduction.  We leave this for a future study.

\vfill\eject
\centerline{\bf References}
\medskip

\item{[1]}
Zamolodchikov, A. B.:
{\it Infinite additional symmetries in 2--dimensional conformal
quantum field theory.}
Theor. Math. Phys. {\bf 65}, 1205-1213 (1985)
\item{[2]}
Lukyanov, S. L., Fateev, V. A.:
{\it Additional symmetries and exactly soluble models in
two--dimensional
conformal field theory.}
Sov. Sci. Rev. {\bf A}. Phys. {\bf 15}, 1-116 (1990)
\item{[3]}
Bouwknegt, P., Schoutens, K.:
{\it ${\cal W}$-symmetry in conformal field theory.}
Phys. Rep. {\bf 223},  183-276 (1993)
\item{[4]}
Blumenhagen, R., Flohr, M., Kliem, A., Nahm, W., Recknagel, A.,
Varnhagen, R.:
{\it $W$-algebras with two and three generators.}
Nucl. Phys. {\bf B361}, 255-289 (1991);
\item{}
Eholzer, W., Honecker, A., H\"ubel, R.:
{\it How complete is the classification of ${\cal W}$-symmetries}?
Phys. Lett. {\bf 308B}, 42-50 (1993)
\item{[5]}
Kausch, H. G., Watts, G. M. T.:
{\it A Study of $W$-algebras using Jacobi identities.}
Nucl. Phys. {\bf B354}, 740-768 (1991)
\item{[6]}
Goddard, P., Kent, A., Olive, D.:
{\it Virasoro algebras and coset space models.}
Phys. Lett. {\bf 152B}, 88-92 (1985)
\item{[7]}
Bais, F. A., Bouwknegt, P., Schoutens, K., Surridge, M.:
{\it Extensions of the Virasoro algebra constructed from
Kac-Moody algebras by using higher order Casimir
invariants.} Nucl. Phys. {\bf B304}, 348-370 (1988);
{\it Coset constructions for extended Virasoro algebras.}
Nucl. Phys. {\bf B304}, 371-391 (1988)
\item{[8]}
Bowcock, P., Goddard, P.:
{\it Coset constructions and extended conformal algebras.}
Nucl. Phys. {\bf B305}, 685-709 (1988)
\item{[9]}
Bouwknegt, P.:
{\it Extended conformal algebras from Kac-Moody algebras.}
In: Infinite dimensional Lie algebras and Lie groups.
Advanced Series in Math. Phys. {\bf 7},
Kac, V. G. (ed.) Singapore:  World Scientific  1989
\item{[10]}
Watts, G. M. T.: {\it $W$-algebras and coset models.}
Phys. Lett. {\bf 245B}, 65-71 (1990)
\item{[11]}
Drinfeld, V. G., Sokolov, V. V.:
{\it Lie algebras and equations of Korteweg - de Vries type.}
Jour. Sov. Math. {\bf 30}, 1975-2036 (1984)
\item{[12]}
Fateev V. A., Lukyanov, S. L.:
{\it The models of two dimensional conformal quantum field
theory with $Z_n$ symmetry.}
Int. J. Mod. Phys. {\bf A3}, 507-520 (1988)
\item{[13]}
Yamagishi, K.:  {\it The KP hierarchy and extended
Virasoro algebras.}
Phys. Lett. {\bf 205B}, 466-470 (1988)
\item{}
Mathieu, P.:
{\it Extended classical conformal algebras and the second Hamiltonian
structure of Lax equations.}
Phys. Lett. {\bf 208B}, 101-106 (1988)
\item{}
Bakas, I.:
{\it The Hamiltonian structure of the spin-4 operator algebra.}
Phys. Lett. {\bf 213B}, 313-318 (1988)
\item{[14]}
Balog, J., Feh\'er, L., Forg\'acs, P., O'Raifeartaigh, L., Wipf, A.:
{\it Toda theory and ${\cal W}$-algebra from a gauged WZNW point
of view.}
Ann. Phys. (N.~Y.) {\bf 203}, 76-136 (1990)
\item{[15]}
Bais, F. A., Tjin, T., van Driel, P.:
{\it Covariantly coupled chiral algebras.}
Nucl. Phys. {\bf B357}, 632-654 (1991)
\item{[16]}
Feh\'er, L., O'Raifeartaigh, L., Ruelle, P., Tsutsui, I., Wipf, A.:
{\it Generalized Toda theories and ${\cal W}$-algebras associated
with integral gradings.} Ann. Phys. (N. Y.) {\bf 213}, 1-20 (1992)
\item{[17]}
Frappat, L., Ragoucy, E., Sorba, P.:
{\it $W$-algebras and superalgebras from constrained WZW models:
a group theoretical classification.}
Lyon preprint ENSLAPP-AL-391/92 (1992)
\item{[18]}
Bershadsky, M, Ooguri, H.:
{\it Hidden $SL(n)$ symmetry in conformal field theories.}
Commun. Math. Phys. {\bf 126}, 49-83 (1989)
\item{[19]}
Figueroa-O'Farrill, J. M.:
{\it On the homological construction of Casimir algebras.}
Nucl. Phys. {\bf B343}, 450-466 (1990)
\item{[20]}
Feigin, B. L., Frenkel, E.:
{\it Quantization of the Drinfeld-Sokolov reduction.}
Phys. Lett. {\bf 246B}, 75-81 (1990)
\item{[21]}
Frenkel, E., Kac, V. G. , Wakimoto, M.:
{\it Characters and fusion rules for $W$-algebras via quantized
Drinfeld-Sokolov reduction.}
Commun. Math. Phys. {\bf 147}, 295-328 (1992)
\item{[22]}
de Boer, J., Tjin, T.:
{\it The relation between quantum $W$ algebras and Lie algebras.}
Utrecht--Amsterdam preprint THU-93/05, IFTA-02-93 (1993),
hep-th/9302006
\item{}
Sevrin, A., Troost, W.:
{\it Extensions of the Virasoro algebra and gauged WZW models.}
preprint LBL-34125, UCB-PTH-93/19, KUL-TF-93/21 (1993),
hep-th/9306033
\item{[23]}
Feh\'er, L., O'Raifeartaigh, L., Ruelle, P., Tsutsui, I., Wipf, A.:
{\it On Hamiltonian reductions of the Wess-Zumino-Novikov-Witten
theories.}
Phys. Rep. {\bf 222}, 1-64 (1992)
\item{[24]}
Bilal, A., Gervais-J.-L.:
{\it Systematic approach to conformal systems with extended
Virasoro symmetries.}
Phys. Lett. {\bf 206B}, 412-420 (1988);
{\it Extended $c=\infty$ conformal systems from classsical Toda
field theories.}
Nucl. Phys. {\bf B314}, 646-686 (1989);
{\it Systematic construction of conformal theories with
higher spin Virasoro symmetries.}
Nucl. Phys. {\bf B318}, 579-630 (1989)
\item{[25]}
Saveliev, M.: {\it On some connections and extensions of $W$-algebras.}
Mod. Phys. Lett. {\bf A5}, 2223-2229 (1990)
\item{[26]}
Mansfield, P., Spence, B.:
{\it Toda theory, the geometry of $\W$-algebras and minimal models.}
Nucl. Phys. {\bf B362}, 294-328 (1991)
\item{[27]}
Dynkin, E. B.:
{\it Semisimple subalgebras of semisimple Lie algebras.}
Amer. Math. Soc. Transl. {\bf 6 [2]}, 111-244 (1957)
\item{[28]}
Polyakov, A. M.:
{\it Gauge transformations and diffeomorphisms.}
Int. J. Mod. Phys. {\bf A5}, 833-842 (1990)
\item{[29]}
Bershadsky, M.: {\it Conformal field theories via Hamiltonian reduction.}
Commun. Math. Phys. {\bf 139}, 71-82 (1991)
\item{[30]}
Feh\'er, L.,  O'Raifeartaigh, L., Ruelle, P.,  Tsutsui, I.:
{\it Rational versus polynomial character of $W_n^l$-algebras.}
 Phys. Lett. {\bf B283},  243-251 (1992)
\item{[31]}
Bowcock, P., Watts, G. M. T.:
{\it On the classification of quantum $W$-algebras.}
Nucl. Phys. {\bf B379}, 63-96 (1992)
\item{[32]}
Goddard, P., Schwimmer, A.:
{\it Factoring out free fermions and superconformal algebras.}
Phys. Lett. {\bf 214B},  209-214 (1988)
\item{[33]}
Feh\'er, L.,  O'Raifeartaigh, L.,   Tsutsui, I.:
{\it The vacuum preserving Lie algebra of a classical $\W$-algebra.}
Bonn--Dublin preprint BONN-HE-93-25, DIAS-STP-93-13 (1993),
hep-th/9307190
\item{[34]}
Beukers, F.:
{\it Differential Galois theory.}
In:  From number theory to physics.
Waldschmidt, M., Moussa, P., Lucke, J.-M., Itzykson, C. (eds.)
Berlin: Springer 1992
\item{[35]}
 Deckmyn, A., Thielmans, K.:
{\it Factoring out free fields.}
preprint KUL-TF-93/26 (1993), hep-th/9306129
\item{[36]}
Delduc, F., Frappat, L., Ragoucy, E., Sorba, P., Toppan F.:
{\it Rational $W$-algebras from composite operators.}
preprint ENSLAPP-AL-429/93, NORDITA-93/47-P (1993)
\item{[37]}
Di Francesco, P., Itzykson, C., Zuber, J.-H.:
{\it Classical $W$-algebras.}
Commun. Math. Phys. {\bf 140}, 543-567 (1991)
\item{[38]}
Bonora, L., Xiong, C. S.:
{\it Covariant $sl_2$ decomposition of the $sl_n$ Drinfeld-Sokolov
equations and the $W_n$-algebras.}
Int. J. Mod. Phys. {\bf A7}, 1507-1525 (1992)

\bye